\documentclass[aps,preprint,epsfig,rotate]{revtex4}
\usepackage{sansmath,amsmath,pdflscape,bm,epsfig,graphicx,rotating}
\usepackage{pdfpages}
\usepackage{pdflscape}
\newcommand{\be}{\begin{equation}}
\newcommand{\ee}{\end{equation}}
\newcommand{\bea}{\begin{eqnarray}}
\newcommand{\eea}{\end{eqnarray}}
\begin{document}

\title{\bf Evaluation of Hylleraas-CI atomic integrals by integration over the coordinates of one electron. 
IV. An improved algorithm for three-electron kinetic energy integrals}
 
\author{Mar\'{\i}a Bel\'{e}n Ruiz}
\email[E--mail address: ]{maria.belen.ruiz@fau.de}

\affiliation{Theoretical Chemistry,\\
Department of Chemistry and Pharmacy, \\
 Friedrich-Alexander-University Erlangen-N\"urnberg, \\
Egerlandstra\ss e 3, 91058 Erlangen, Germany}

\date{\today}

\begin{abstract}
\noindent An improved algorithm to evaluate the nonrelativistic three-electron 
Hylleraas-Configuration Interaction (Hy-CI) kinetic energy integrals over Slater 
orbitals and the Hamiltonian in Hylleraas coordinates is shown. 
The resulting analytical expressions are general for all quantum numbers of the orbitals. 
From there, the restriction of employing orbitals with quantum numbers $l \le 2$ of the 
above algorithm presented in paper I of this series has been removed. With the new algorithm  
it is possible, in the direct integration method described in this series, 
to carry out Hy-CI atomic structure calculations including $f$-, $g$-, \ldots , $l$  
and higher angular-momentum Slater orbitals and to determine $F$, $G$, \ldots , $L$ and states of 
higher order symmetry. 

\end{abstract}


\maketitle

\newpage

\section{Introduction}

Hylleraas-Configuration Interaction (Hy-CI) wave functions \cite{Hy-CI1,Hy-CI2} are 
of great importance in Quantum Chemistry because being general for any atom, lead to 
high precision energy values of atomic levels and various properties.   

The kinetic energy integrals, which are generated by the kinetic energy
operator part of the Hamiltonian \cite{Ruiz1}, are not as complex as the repulsion
four-electron integrals, but their evaluation is also difficult. For this reason, they need 
a separate treatment.  
There are two kinds of kinetic energy integrals in the Hy-CI method: the two-electron 
kinetic energy integrals, needed not only in the computation of two-electron systems, 
but also generally in calculations of any N-electron system; and the 
three-electron kinetic energy integrals, which occur in the computation of three-electron 
and larger systems.  

In a recent paper III of these series \cite{Ruiz2e},
we have evaluated the two-electron kinetic energy integrals by two methods: 1) the
so-called Kolos and Roothaan transformation (KR) \cite{KR}; and 2) the method of the 
direct application of the differenciating operators on the wave function \cite{Ruiz3e}, 
obtaining completely agreement in the values of the integrals by both methods
(more than 30 decimal digits in our computer using quadruple precision). 
For the two-electron kinetic energy integrals, the KR  
transformation has proven to be computationally faster than the direct differentiation 
\cite{Ruiz3e}, with identical memory requirements.
Unfortunately, a similar transformation as KR in the three-electron case has not been yet achieved. 
Therefore in the case of the three-electron kinetic energy integrals we have to use 
the method of direct differentiation of \cite{Ruiz3e}. 
This includes the performance of spherical harmonics derivatives 
and the use of recursion relations over these functions.

During the mathematical evaluation the major difficulty encountered consisted in treating
the recursion relation involving the inverse sine function \cite[Eq. (C.6)]{Ruiz3e}: 

\begin{gather}
-\frac{Y_L^M(\theta _1,\varphi _1)}{\sin \theta _1}=\frac 1{2M}\left[ \frac{%
(2L+1)}{(2L+3)}\right] ^{1/2}\Big[ [(L-M+1)(L-M+2)]^{1/2}
Y_{L+1}^{M-1}(\theta _1,\varphi _1) e^{i\varphi_1} \nonumber \\
 +\left[(L+M+1)(L+M+2)\right]^{1/2} Y_{L+1}^{M+1}(\theta _1,\varphi_1)
e^{-i\varphi _1} \Big], \qquad M\ne 0,  
\end{gather}

\noindent since its use is conditioned by $M \ne 0$. 
Not being aware of any other recursion relation containing the inverse of the sine function, in 
order to avoid singularities, we distinguished in paper I among the cases $M=0$, $M>0$, and $M<0$. 
While for the case $M=0$ a 
general formula could be found \cite[Eq. (C.9)]{Ruiz3e}, for the cases $M>0$ and $M<0$ the   
expressions were too lengthly. So we were able to found recursion relations, 
which were restricted to the cases $L \le 2$ \cite[Eq. (C.10, C.11)]{Ruiz3e}. Although their 
generalization is possible, it seems more reasonable to look for a more efficient procedure.   

In this work we will use Eq. (1) such, that the condition $M\ne 0$ will be always 
fulfilled and we will obtain general expressions for the integral. 
All types kinetic energy integrals will be revised and 
new compact expressions will be presented. Finally, computed values of selected integrals of Refs. \cite{Ruiz3e,Sims3e} 
and of new kinetic energy integrals are given in this work. 

The obtained relations have been programmed in Fortran 90 language and the resulting 
integral values have been compared with the ones obtained by the older algorithm of paper I, 
finding complete agreement. The new subroutines have replaced the previous ones in our Hy-CI 
computer code and calculations have been done for 
S, P, D, F, G, H, K, and L states of the Li atom, showing that the algorithm performs properly and 
it is stable. Moreover, we have checked the final printed formulas of this article with the computer  
Fortran code.    

Before to start with the alternative procedure to evaluate the kinetic energy integrals, 
let us say some general words about the two methods of evaluation of Hy-CI integrals. 
In this series of papers I-IV \cite{Ruiz2e,Ruiz3e,Ruiz4e} we use the method of direct 
integration over the interelectronic distance $r_{ij}$ and the coordinates of one of 
the electrons, reducing integrals to new ones with lower number of electrons. 
The three- and four-electron 
integrals are then reduced to a linear combination of basic two-electron
integrals and these ones are evaluated as a sum of two-electron auxiliary integrals. 
This method of integration has computational advantages in storage memory 
and facilitates extension to larger systems. 

The earlier method is the one of Sims and Hagstrom \cite{Sims3e,Sims4e}, where 
the interelectronic distances $r_{ij}$ are expanded into one-electron distances. 
The expansion of the
interelectronic distances is a concept they use systematically in the
evaluation of all kind of integrals. This fact leads to the appearance of
three-electron $W$ and four-electron $X$ auxiliary integrals. 
On the contrary, when using our method for 
the three-electron kinetic energy integrals, no three-electron auxiliary
integrals are needed, with a great saving of computer memory. Note that the
auxiliary $W$ integrals are three-fold and have to be calculated for a large
number of powers and exponents. The number of exponents grows with the
atomic number.

\section{Kinetic energy integrals}

Let us define the Slater orbitals of electron $i$ specified by the quantum numbers $n_i$, $m_i$ and 
$l_i$ and orbital exponent $\alpha_i$ with an unnormalized radial part and orthonormal spherical harmonics: 
\begin{eqnarray}
\phi^{*}(\mathbf{r}_i) &=&r_i^{n_i-1}e^{-\alpha_i r_i}Y_{l_i}^{m_i^*}(\theta_i ,\varphi_i ), 
\nonumber \\
\phi^{\prime }(\mathbf{r}_i) &=&r_i^{n_i^{\prime }-1}e^{-\alpha_i^{\prime
}r_i}Y_{{l_i^{\prime }}}^{m_i^{\prime }}(\theta_i ,\varphi_i ).
\end{eqnarray}
The symbol ${}^*$ means the complex conjugate. 
The spherical harmonics in Condon and Shortley phases \cite[p. 52]{Condon}
are given by: 
\begin{equation}
Y_{l_i}^{m_i}(\theta_i ,\varphi_i )=(-1)^{m_i}\left[ \frac{(2l_i+1)}{4\pi }\frac{(l_i-m_i)!}{(l_i+m_i)!}%
\right]^{1/2}P_{l_i}^{m_i}(\cos {\theta_i })e^{im_i\varphi_i },
\end{equation}
with the associated Legendre functions $P_{l_i}^{m_i}(\cos {\theta_i }).$ The
spherical harmonics and associated Legendre functions used along this work
are written explicitly in \cite[p. 14]{Stevenson}, and defined as in Ref. 
\cite{Hy-CI1}. They obey the condition: 
\begin{equation}
Y_{l_i}^{m_i^*}(\theta_i ,\varphi_i )=(-1)^{m_i}Y_{l_i}^{-m_i}(\theta_i ,\varphi_i ).
\end{equation}

We define the one-electron charge distributions by expanding or linearizing
the products of spherical harmonics with equal argument $i$ using the formula 
\cite[Eq. (12)]{Hy-CI1}: 
\begin{equation}
Y_{l_i}^{m_i^*}(\theta_i ,\varphi_i )Y_{{l_i}^{\prime }}^{m_i^{\prime }}(\theta_i ,\varphi_i
)=\sum_{L_i=|l_i-l_i^{\prime }|}^{l_i+l_i^{\prime }}{}\left[ \frac{2L_i+1}{4\pi }\right]
^{1/2}C^{L_i}(l_i^{\prime },m_i^{\prime };l_i,m_i)Y_{L_i}^{m_i^{\prime }-m_i}(\theta_i ,\varphi_i ),
\end{equation}
$L_i$ satisfies the triangular condition $|l_i-l_i^{\prime }|\leq L_i\leq
l_i+l_i^{\prime }$ and the restriction $L_i\geq |M_i|$. The summation is
done in steps of two: $L_i=|l_i-l_i^{\prime }|,|l_i-l_i^{\prime }|+2,\ldots
,l_i+l_i^{\prime }-2,l_i+l_i^{\prime }$ and $M_i=m_i^{\prime }-m_i$. The
lowest value of $L_i$ depends also on $M_i$. 

The Condon-Shortley coefficients \cite[Eqs. (6-11)]{Condon} are
defined by: 
\begin{equation}
C^{L_i}(l_i^{\prime },m_i^{\prime };l_i,m_i)=\left[ \frac{4\pi }{2L_i+1}\right]^{1/2} \int
Y_{L_i}^{m_i^{\prime }-m_i}(\theta_i ,\phi_i )Y_{{l_i}^{\prime }}^{m_i^{\prime }*}(\theta_i
,\varphi_i )Y_{l_i}^{m_i}(\theta_i ,\varphi_i )\sin {\theta_i }d\theta_i d\varphi_i . 
\end{equation}

We define $N_i=n_i+n_i^{\prime }-1$, and the exponents $\omega_i =\alpha_i +\alpha_i
^{\prime }$. In the next we will use uppercase letters $N_i$, $L_i$, $M_i$ for the quantum
numbers of charge distributions, while lowercase letters $n_i,l_i,m_i$ will be used for
the quantum numbers of the orbitals.

For a $n_e$-electron system the kinetic energy operator in Hylleraas coordinates
can be written \cite{Ruiz1}:
\begin{multline}
\hat{T} =-\frac 12\sum_{i=1}^{n_{e}}\frac{\partial ^2}{\partial r_i^2}%
-\sum_{i=1}^n\frac 1{r_i}\frac \partial {\partial r_i}
-\sum_{i<j}^n\frac{\partial ^2}{\partial r_{ij}^2}-\sum_{i<j}^n\frac
2{r_{ij}}\frac \partial {\partial r_{ij}}  \\
-\frac 12\sum_{i\neq j}^n\frac{r_i^2+r_{ij}^2-r_j^2}{r_ir_{ij}}\frac{%
\partial ^2}{\partial r_i\partial r_{ij}}-\frac 12\sum_{i\neq
j}^n\sum_{k>j}^n\frac{r_{ij}^2+r_{ik}^2-r_{jk}^2}{r_{ij}r_{ik}}\frac{%
\partial ^2}{\partial r_{ij}\partial r_{ik}}  \\
-\frac 12\sum_{i=1}^n\frac 1{r_i^2}\frac{\partial ^2}{\partial \theta _i^2}%
-\frac 12\sum_{i=1}^n\frac{\cot {\theta _i}}{r_i^2}%
\frac \partial {\partial \theta _i} 
-\frac 12\sum_{i=1}^n\frac 1{r_i^2\sin ^2{\theta _i}}\frac{\partial ^2}{%
\partial \varphi _i^2}  \\
-\sum_{i\neq j}^n\left( \frac{r_j}{r_ir_{ij}}\frac{\cos {\theta _j}}{\sin {%
\theta _i}}+\frac 12\cot {\theta _i}\frac{r_{ij}^2-r_i^2-r_j^2}{r_i^2r_{ij}}%
\right) \frac{\partial ^2}{\partial \theta _i\partial r_{ij}}  \\
-\sum_{i\neq j}^n\frac{r_j}{r_ir_{ij}}\frac{\sin {\theta _j}}{\sin {\theta
_i}}\sin {(\varphi _i-\varphi _j)}\frac{\partial ^2}{\partial \varphi
_i\partial r_{ij}}\ .
\end{multline}

When $\hat T$ is applied to a wave function containing at most one $r_{ij}$ per configuration
all terms arising from $\frac{\partial ^2} {\partial r_{ij}^2}$ and $\frac{\partial ^2} {%
\partial r_{ij}\partial r_{ik}}$ vanish.

The one-electron angular-momentum operators can be identified and replace by its eigenvalues:
\begin{equation}
\hat{L_i^2}=-\frac{\partial^2}{\partial \theta_i^2}
-\cot{\theta_i}\frac \partial {\partial \theta_i}
-\frac1{\sin^2{\theta_i}}\frac{\partial^2}{\partial \varphi_i^2},
\end{equation}
where
\begin{equation}
\hat{L_i^2} Y_{l_i}^{m_i}(\theta_i ,\varphi_i ) = l_i(l_i+1) Y_{l_i}^{m_i}(\theta_i ,\varphi_i ). 
\end{equation}

The kinetic energy operator $\hat T$ of Eq. (7) can be separated into kinetic energy operators acting on every 
electron $i$, which can be again separated into radial and angular 
parts. As a pattern integral, let us evaluate in this work the kinetic energy of electron $1$, i.e. $T(1)$, 
of a term containing the interelectronic distance $r_{12}$ on the right-hand side and $r_{13}$ on the left-hand side 
of a matrix element. Then $T(1)$ is built up with the following contributions:  

\begin{eqnarray}
\hat{T}(1)&=& \hat{T}_{R}(1)+\hat{T}_{\theta,\varphi}(1), \nonumber \\ 
\hat{T}_{\theta,\varphi}(1) &=& \hat{T}_L(1) + \hat{T}_{\theta ,1}(1) + \hat{T}_{\theta ,2}(1) 
+ \hat{T}_{\varphi}(1). 
\end{eqnarray}

\noindent The radial parts do not present any difficulty and have been already 
evaluated in our previous work \cite{Ruiz3e}. The angular parts of the kinetic energy operator are: 
 
\begin{eqnarray}
\hat{T}_L(1) &=&\frac 12\frac{\hat{L}^2(1)}{r_1^2}, \\
\hat{T}_{\theta ,1}(1) &=&-\frac{r_2}{r_1r_{12}}\frac{\cos {\theta _2}}{\sin 
{\theta _1}}\frac{\partial ^2}{\partial \theta _1\partial r_{12}}, \\
\hat{T}_{\theta ,2}(1) &=&-\frac 12\cot {\theta _1}\frac{r_{12}^2-r_1^2-r_2^2%
}{r_1^2r_{12}}\frac{\partial ^2}{\partial \theta _1\partial r_{12}}, \\
\hat{T}_{\varphi }(1) &=&-\frac{r_2}{r_1r_{12}}\frac{\sin {\theta _2}}{\sin 
{\theta _1}}\sin {(\varphi_1-\varphi_2)}\frac{\partial ^2}{\partial \varphi
_1\partial r_{12}}\ . 
\end{eqnarray}

The expectation value of the angular momentum operator is evaluated 
using the eigenvalue equation Eq. (9).  

The evaluation of the angular kinetic energy contributions from the operators 
$\hat{T}_{\theta ,1}(1)$, $\hat{T}_{\theta ,2}(1)$, and $\hat{T}_{\varphi }(1)$ is more
involved and is reported in detail in the Appendices A, B, and C of this work. 
In this Section we are presenting the final expressions. The first one is the  
kinetic energy integral $I_{\theta ,1}(1)$, i.e. for electron 
$1$, corresponding to the operator $\hat{T}_{\theta ,1}(1)$ can be computed with the following 
equation:  

\begin{multline}
I_{\theta ,1}(1)=\left\langle \phi (\mathbf{r}_1)\phi (\mathbf{r}_2)\phi (\mathbf{r}%
_3)r_{13}|\hat{T}_{\theta ,1}(1) |\phi (\mathbf{r}%
_1)\phi (\mathbf{r}_2)\phi (\mathbf{r}_3)r_{12}\right\rangle = \delta (M_1+M_2+M_3,0) \\  
\times \sum_{L_1=|l_1^{\prime }-l_1|}^{l_1^{\prime
}+l_1}{}\sum_{L_3=|l_3^{\prime }-l_3|}^{l_3^{\prime }+l_3}C_{L_3} \Bigg\{ 
\sum_{L_2=|l_2^{\prime }-1-l_2|}^{l_2^{\prime
}-1+l_2}{}\sum_{K_1=|L_2-L_1|}^{L_1+L_2}\sum_{K_2=|K_1-L_3|}^{K_1+L_3} F_1 \; C_{L_2a}f_{2a}\\ 
\times \Big(f_{1a}f_{a}(K_2,M)\;C_{L_1a}C_{K_1a}C_{K_2a}\;B_{1}(K_2+1,M+1) \\
-f_{1b}f_{b}(K_2,M^{\prime})\;C_{L_1b}C_{K_1b}C_{K_2b}\;B_{1}(K_2+1,M^{\prime}-1)\Big)  \\
\times J(N_1-1,N_3,N_2+1;\omega _1,\omega _3,\omega _2;1,-1;L_3,L_2) \\ 
+\sum_{L_2^{\prime }=|l_2^{\prime }+1-l_2|}^{l_2^{\prime
}+1+l_2}{}\sum_{K_1^{\prime }=|L_2^{\prime }-L_1|}^{L_1+L_2^{\prime}}\sum_{K_2^{\prime
}=|K_1^{\prime }-L_3|}^{K_1^{\prime }+L_3} F_2 \; C_{L_2b} f_{2b} \\ 
\times \Big( f_{1a}f_{a}(K_2^{\prime},M)\;C_{L_1a}C_{K_1c}C_{K_2c}\;B_{1}(K_2^{\prime }+1,M+1)  \\
 -f_{1b}f_{b}(K_2^{\prime},M^{\prime})\;C_{L_1b}C_{K_1d}C_{K_2d}\;
B_{1}(K_2^{\prime }+1,M^{\prime }-1)\Big)  \\
  \times J(N_1-1,N_3,N_2+1;\omega _1,\omega _3,\omega_2;1,-1;L_3,L_2^{\prime })\Bigg\}. 
\end{multline}

\noindent This expression is a limited sum of radial three-electron integrals $J$'s (the terms in which 
three-electron integrals are expanded), some factors $f$'s and 
Condon and Shortley coefficients $C$'s. The angular factors $f$'s are simple expressions 
containing the quantum numbers of the orbitals or linear combinations of them: 

\begin{eqnarray}
f_{1a}=\frac 12[(l_1^{\prime }+m_1^{\prime }+1)(l_1^{\prime }-m_1^{\prime
})]^{1/2}, & \qquad & f_{1b}=\frac 12[(l_1^{\prime }-m_1^{\prime }+1)(l_1^{\prime
}+m_1^{\prime })]^{1/2}, \nonumber \\
f_{2a}=\left[ \frac{(l_2^{\prime }+m_2^{\prime })(l_2^{\prime }-m_2^{\prime
})}{(2l_2^{\prime }+1)(2l_2^{\prime }-1)}\right] ^{1/2}, &\qquad & f_{2b}=\left[ 
\frac{(l_2^{\prime }+m_2^{\prime }+1)(l_2^{\prime }-m_2^{\prime }+1)}{%
(2l_2^{\prime }+1)(2l_2^{\prime }+3)}\right] ^{1/2}, \nonumber \\
f_a(K,M)=\frac 1{2M}[(K+M+2)(K+M+1)]^{1/2}, &\qquad & 
f_b(K,M)=\frac 1{2M}[(K-M+2)(K-M+1)]^{1/2}. \nonumber \\
\end{eqnarray}
$M=M_1+M_2+M_3+1$ and $M^{\prime }=M_1+M_2+M_3-1$. The indices $K$'s are linear combinations of $L$'s. 
There are some general factors too: 
\begin{eqnarray}
F_1 &=& \frac{(-1)^{M_2+M_3}}{\sqrt{4\pi}}\Bigg[\frac{(2K_2+1)}{(2K_2+3)}\Bigg]^{1/2} 
\Big[(2L_1+1)(2L_2+1)(2L_3+1)(2K_1+1)(2K_2+1)\Big]^{1/2}, \nonumber \\ 
F_2 &=& \frac{(-1)^{M_2+M_3}}{\sqrt{4\pi}}\Bigg[\frac{(2K_2^{\prime}+1)}{(2K_2^{\prime}+3)}\Bigg]^{1/2} 
\Big[ (2L_1+1)(2L_2^{\prime }+1)(2L_3+1)(2K_1^{\prime }+1)(2K_2^{\prime}+1)\Big] ^{1/2}, \nonumber \\ 
\end{eqnarray}
the $C$'s are the Condon and Shortley coefficients listed here: 

\begin{eqnarray}
C_{L_1a}=C^{L_1}(l_1^{\prime },m_1^{\prime }+1;l_1,m_1), &\qquad & 
C_{L_1b}=C^{L_1}(l_1^{\prime },m_1^{\prime }-1;l_1,m_1), \nonumber \\ 
C_{L_2a}=C^{L_2}(l_2^{\prime }-1,m_2^{\prime };l_2,m_2), &\qquad & 
C_{L_2b}=C^{L_2^{\prime }}(l_2^{\prime }+1,m_2^{\prime };l_2,m_2), \nonumber \\
C_{L_3}=C^{L_3}(l_3^{\prime },m_3^{\prime };l_3,m_3), & &  \nonumber \\
C_{K_1a}=C^{K_1}(L_1,M_1+1;L_2,-M_2),&\qquad & C_{K_1b}=C^{K_1}(L_1,M_1-1;L_2,-M_2), \nonumber \\  
C_{K_1c}=C^{K_1^{\prime }}(L_1,M_1+1;L_2^{\prime },-M_2), &\qquad & 
C_{K_1d}=C^{K_1^{\prime }}(L_1,M_1-1;L_2^{\prime },-M_2), \nonumber \\ 
C_{K_2a}=C^{K_2}(K_1,M_1+M_2+1;L_3,-M_3),&\qquad& C_{K_2b}=C^{K_2}(K_1,M_1+M_2-1;L_3,-M_3),  \nonumber \\
C_{K_2c}=C^{K_2^{\prime }}(K_1^{\prime },M_1+M_2+1;L_3,-M_3), &\qquad & 
C_{K_2d}=C^{K_2^{\prime }}(K_1^{\prime },M_1+M_2-1;L_3,-M_3), \nonumber \\
\end{eqnarray}

\noindent and $B_1$ are special one-electron angular integrals over spherical harmonics 
which can be calculated using the algorithm of Wong \cite{Wong}: 

\begin{equation}
B_{1}(L,M+n)=\pi ^{1/2}(2L+1)^{1/2}\left[ \frac{(L-M-n)!}{(L+M+n)!}\right]
^{1/2} P(L,M+n), 
\end{equation}

\noindent with
\begin{equation}
P(L,M+n)=\sum_{p=0}^{p_{\max }} a_{L,M}^{p} 
\frac{\Gamma \left( \frac 12(L-M-2p+1)\right)
\Gamma \left( \frac 12(M+2p+2)\right) }{\Gamma \left( \frac
12(L+3)\right) }, 
\end{equation}
   
\noindent with coefficients:
\begin{equation}
a_{L_{,}M}^p=\frac{(-1)^p(L+M)!}{2^{M+2p}(M+p)!p!(L-M-2p)!}.
\end{equation}

\bigskip

\noindent $p_{\max }=[(L-M)/2]$ is the integral part of
$(L-M)/2$ and $\Gamma $ are Gamma functions. For more details, see Appendix A. Finally, 
the radial three-electron integrals $J$ \cite{Ruiz3e} can be computed using the following 
formula:  

\begin{multline}
J(N_{1},N_{2},N_{3};\omega _{1},\omega _{2},\omega _{3};1,-1 ;L_{2},L_{3})=
\\
\sum_{k=0}^{\lfloor L_{2}/2\rfloor
}\sum_{q=0}^{L_{2}-2k}\sum_{p=0}^{L_{2}-2k-q}\frac{(-1)^{k+q}}{%
2^{2L_{2}-2k}(2q+3)}{\binom{L_{2}}{k}\binom{2L_{2}-2k}{L_{2}}\binom{L_{2}-2k%
}{q}}{\binom{L_{2}-2k-q}{p}} \\
\times \left\{ \sum_{i=1}^{q+2}{\binom{2q+3}{2i-1}}A(N_{2}-1+2k+2p+2i-L_{2},%
\omega _{2})\right. \text{ } \\
\times I(N_{1}+L_{2}+3-2k-2p-2i,N_{3};\omega _{1},\omega _{3};-1 ;L_{3}) \\
-\sum_{j=1}^{N_{2}+2k+2p+1-L_{2}}{\binom{N_{2}+2k+2p-L_{2}}{j-1}}%
A(2q+2+j;\omega _{2}) \\
\left. \times \text{ }I(N_{1}+N_{2}-2q-j,N_{3};\omega _{1}+\omega
_{2},\omega _{3};-1 ;L_{3})\right. \Bigg\} . 
\end{multline} 

The $J$'s are limited sums of basic radial two-electron integrals $I$'s, which are computed 
with high accuracy (about 30 decimal digits in our computer):

\begin{equation}
I(N_1,N_2;\omega _1,\omega _2;-1;L)=\frac 1{(2L+1)} \Big[ 
V(N_1+L+1,N_2-L;\omega _1,\omega _2) \nonumber \\
+ V(N_2+L+1,N_1-L;\omega _2,\omega _1) \Big]. 
\end{equation}

\noindent These integrals consist of the sum of two auxiliary integrals.
The two-electron auxiliary integrals $V(m,n;\alpha ,\beta )$ and one-electron
auxiliary integrals $A(n,\alpha )$ have been extensively discussed in papers 
I, II, and III of this series. $V$  integrals are once
calculated with high precision for all different pairs of orbital exponents which may 
occur in the calculation and then stored.

\newpage 


The second angular kinetic energy integral, see Appendix B, is: 

\begin{multline}
I_{\theta ,2}(1)=\left\langle \phi (\mathbf{r}_1)\phi (\mathbf{r}%
_2)\phi (\mathbf{r}_3)r_{13}|\hat T_{\theta ,2}|\phi^{\prime } (%
\mathbf{r}_1)\phi^{\prime } (\mathbf{r}_2)\phi^{\prime } (\mathbf{r}_3)r_{12}\right\rangle =  \delta (M_1+M_2+M_3,0)\\
\times \sum_{L_1=|l_1^{\prime
}-l_1|}^{l_1^{\prime }+l_1}{}{}\sum_{L_2=|l_2^{\prime }-l_2|}^{l_2^{\prime
}+l_2}{}\sum_{L_3=|l_3^{\prime }-l_3|}^{l_3^{\prime
}+l_3}\sum_{L=|L_1-L_2|}^{L_1+L_2}\sum_{L^{\prime }=|L-L_3|}^{L+L_3} F\; C_{L_2}C_{L_3} \\
\times \Bigg\{ f_{1a}f_{2a}C_{1a}C_{La}C_{Lc}\;B_{1}(L^{\prime},M+1) 
- f_{1b}f_{2d}C_{1b}C_{Lb}C_{Ld}\; B_{1}(L^{\prime },M-1) \\ 
+\delta (L^{\prime },0) \Big(
f_{1a}f_{2b}C_{1a}C_{La}C_{Lc}-f_{1b}f_{2c}C_{1b}C_{Lb}C_{Ld}\Big) \Bigg\} \\
\times \Big[ J(N_1-2,N_2,N_3;\omega _1,\omega _2,\omega_3;1,1;L_2,L_3) 
-J(N_1,N_2;N_3;\omega _1,\omega _2,\omega _3;-1,1;L_2,L_3) \\
-J(N_1-2,N_2+2,N_3;\omega _1,\omega _2,\omega _3;-1,1;L_2,L_3)\Big],  
\end{multline}

\bigskip
\noindent with $M=M_1+M_2+M_3+1$ and $M^{\prime }=M_1+M_2+M_3-1$. The factors are:

\begin{eqnarray}
f_{1a}=\frac 12[(l_1^{\prime }+m_1^{\prime }+1)(l_1^{\prime }-m_1^{\prime
})]^{1/2}, & \qquad & 
f_{1b}=\frac 12[(l_1^{\prime }-m_1^{\prime }+1)(l_1^{\prime
}+m_1^{\prime })]^{1/2}, \nonumber \\ 
f_{2a}=\frac 1{2M}[(L^{\prime }+M+1)(L^{\prime }-M)]^{1/2}, & \qquad & 
f_{2b}=\frac 1{2M}[(L^{\prime }-M+1)(L^{\prime }+M)]^{1/2}, \nonumber \\
f_{2c}=\frac 1{2M^{\prime }}[(L^{\prime }+M^{\prime }+1)(L^{\prime
}-M^{\prime })]^{1/2}, & \qquad & 
f_{2d}=\frac 1{2M^{\prime }}[(L^{\prime }+M^{\prime }+1)(L^{\prime
}+M^{\prime })]^{1/2}, \nonumber \\ 
\end{eqnarray}

\noindent a general factor:

\begin{equation}
F=\frac{(-1)^{M_2+M_3}}{4\sqrt{\pi}}\Big[ (2L_1+1)(2L_2+1)(2L_3+1)(2L+1)(2L^{\prime }+1)\Big]^{1/2}, 
\end{equation}

\noindent and the coeficients: 
\begin{eqnarray}
C_{L_1a}=C^{L_1}(l_1^{\prime },m_1^{\prime }+1;l_1,m_1), &\qquad &
C_{L_1b}=C^{L_1}(l_1^{\prime },m_1^{\prime }-1;l_1,m_1), \nonumber \\ 
C_{L_2}=C^{L_2}(l_2^{\prime },m_2^{\prime };l_2,m_2), & \qquad & 
C_{L_3}=C^{L_3}(l_3^{\prime },m_3^{\prime };l_3,m_3),  \nonumber \\
C_{L_a} =C^L(L_1,M_1+1;L_2,-M_2),& \qquad & C_{L_b}=C^L(L_1,M_1-1;L_2,-M_2), \nonumber \\
C_{L_c} =C^{L^{\prime }}(L_1,M_1+1;L_3,-M_3),& \qquad & C_{L_d}=C^{L^{\prime
}}(L_1,M_1-1;L_3,-M_3). 
\end{eqnarray}

Again this integral consists on a limited sum of radial three-electron integrals and 
it can computed with high accuracy. 

\noindent Finally, the third integral, see Appendix C, vanishes if $m_1^{\prime }=0$,  
while for $m_1^{\prime } \neq 0$ can be evaluated using the following programmable 
expression:  

\begin{multline}
I_{\varphi}(1)= \left\langle \phi (\mathbf{r}_1)\phi (\mathbf{r}%
_2)\phi (\mathbf{r}_3)r_{13}|\hat T_{\varphi}|\phi^{\prime } (%
\mathbf{r}_1)\phi^{\prime } (\mathbf{r}_2)\phi^{\prime } (\mathbf{r}_3)r_{12}\right\rangle = \delta (M_1+M_2+M_3,0) \\
\times \sum_{L_3=|l_3^{\prime }-l_3|}^{l_3^{\prime
}+l_3}\sum_{L_2=|l_2-1|}^{l_2+1}\sum_{L_2^{\prime }=|L_2-l_2^{\prime
}|}^{L_2+l_2^{\prime }}\sum_{L_1=|l_1^{\prime }+1-l_1|}^{l_1^{\prime }+1+l_1}
\sum_{L_3^{\prime }=|L_3-L_2^{\prime }|}^{L_3+L_2^{\prime }} F \;C_{L_3} \\ 
\times \Bigg\{ C_{L_2a}C_{L_2^{\prime }a}
\Big( f_{1a}C_{L_1a}C_{L_3^{\prime}a}B_{2}(L_1,M_1-1;L_3^{\prime },M_2+M_3-1) 
+ (-1)^{M_1+1}\delta (L_1,L_3^{\prime })f_{1b}C_{L_1b}C_{L_3^{\prime}b} \Big) \\
+C_{L_2b}C_{L_2^{\prime }b}\Big(
f_{1b}C_{L_1b}C_{L_3^{\prime}b}B_{2}(L_1,M_1+1;L_2^{\prime },M_2+M_3+1)
+(-1)^{M_1-1}\delta(L_1,L_3^{\prime }) f_{1a}C_{L_1a}C_{L_3^{\prime}a} 
\Big) \Bigg\} \\ 
\times J(N_1-1,N_3,N_2+1;\omega _1,\omega _3,\omega_2;1,-1;L_3,L_2^{\prime }).
\end{multline}

The factors are defined: 
\begin{eqnarray}
f_{1a}=\Big[(l_1^{\prime }-m_1^{\prime }+2)(l_1^{\prime }-m_1^{\prime
}+1)\Big]^{1/2}, &\qquad&  f_{1b}=\Big[(l_1^{\prime }+m_1^{\prime }+2)(l_1^{\prime
}+m_1^{\prime }+1)\Big]^{1/2}, 
\end{eqnarray}
a general factor:

\begin{equation}
F = \frac{(-1)^{m_2^{\prime }+M_3}\sqrt{2}}{4\sqrt{3}} \left[ \frac{2l_1^{\prime }+1}{2l_1^{\prime }+3}\right]^{1/2}
\Big[(2L_1+1)(2L_2+1)(2L_3+1)(2L_2^{\prime }+1)(2L_3^{\prime }+1)\Big]^{1/2}, 
\end{equation}

\noindent and the coefficents: 
\begin{eqnarray}
C_{L_1a}=C^{L_1}(l_1^{\prime }+1,m_1^{\prime }-1;l_1,m_1),& \qquad & 
C_{L_1b}=C^{L_1}(l_1^{\prime }+1,m_1^{\prime }+1;l_1,m_1), \nonumber \\
C_{L_2a}=C^{L_2}(1,-1;l_2,m_2),& \qquad & C_{L_2b}=C^{L_2}(1,1;l_2,m_2), \nonumber \\
C_{L_2^{\prime }a}=C^{L_2^{\prime }}(L_2,-1-m_2;l_2^{\prime },-m_2^{\prime}), &\qquad& 
C_{L_2^{\prime }b}=C^{L_2^{\prime }}(L_2,1-m_2;l_2^{\prime },-m_2^{\prime }), \nonumber \\
C_{L_3}=C^{L_3}(l_3^{\prime },m_3^{\prime };l_3,m_3), & & \nonumber \\
C_{L_3^{\prime}a}=C^{L_3^{\prime }}(L_2^{\prime },M_2+1;L_3,-M_3), &\qquad &
C_{L_3^{\prime}b}=C^{L_3^{\prime }}(L_2^{\prime },M_2-1;L_3,-M_3).
\end{eqnarray}

A new auxiliary integral is used: 
\begin{multline}
B_{2}(L_1,M_1+n;L_2,M_2+n)=\frac 12\Big[ (2L_1+1)(2L_2+1)\Big]^{1/2}
\left[ \frac{(L_1-M_1-n)!}{(L_1+M_1+n)!}\right]^{1/2} \left[ \frac{(L_2-M_2-n)!}{%
(L_2+M_2+n)!}\right]^{1/2} \\
\times \sum_{p_{1=0}}^{p_{1\max }}\sum_{p_{2=0}}^{p_{2\max
}}a_{L_{1,}M_1}^{p_1}a_{L_{2,}M_2}^{p_2} 
\frac{\Gamma \left( \frac 12(L_1+L_2-M_1-M_2-2p_1-2p_2+1)\right)
\Gamma \left( \frac 12(M_1+M_2+2p_1+2p_2+2)\right) }{\Gamma \left( \frac
12(L_1+L_2+3)\right) }.
\end{multline}
The coefficients $a_{L,M}^p$ are given in Eq. (20). Further details can be found in Appendix C. 

The expressions shown above shown have been programmed into a computer code and results values of the three types 
of kinetic energy integrals obtained. The results are shown in Table I. There, it can be found not only the 
total value of the kinetic energy integral but also the contributions $I_R$, $I_L$, $I_{\theta ,1}$, $I_{\theta ,2}$, 
and $I_{\varphi}$ corresponding to the final expressions Eqs. (15,23,27). 
The radial contribution $I_R$ was already discussed 
in paper I of this series \cite{Ruiz3e}. Some of the integrals of Table I are the same ones than in Table 6 of Ref. 
\cite{Ruiz3e}, some are reproduced values of integrals from Table 1 of the paper of Sims and Hagstrom \cite{Sims3e}, 
and the last values are new integrals obtained with non-vanishing $I_{\varphi}$ contributions, which represent 
rare cases.  
The agreement with previous values of paper I of this series and the ones of Sims and Hagstrom is complete (about 30 
decimal digits in our computer using quadruple precision).


\begin{center}
\begin{table}[tbp]
\caption{\footnotesize Values and partitioning of the kinetic energy three-electron integrals of electron 1 of 
Table 6 of Ref. \cite{Ruiz3e} and several integrals of Table 1 of Ref. \cite{Sims3e}. The charge distributions are constructed with the exponents $\omega_i$=1.40 for
orbitals with ${}^{\prime \prime }$, otherwise $\omega_i=2.86$. The sum of the contributions to the total kinetic 
energy integral $I_R$, $I_L$, $I_{\theta ,1}$, $I_{\theta ,2}$, and $I_{\varphi}$ leads to the total value listed 
above them. Non-printed contributions are zero.} 
\scalebox{0.75}{ 
\begin{tabular}{ccccc}
\hline\hline
Charge distribution & $\omega_1 $ & $\omega_2 $ & $\omega_3$ & $I_{KE}$ \\
\hline
$(1s1s,1s1s^{\prime \prime },1s1s^{\prime \prime })$ & 5.72 & 4.26 & 4.26 & 
$\;$0.15659 17112 60607 62842 78921 37447 $\times 10^{-4}$ \\
$I_R$ & & & &         $\;$0.15659 17112 60607 62842 78921 37447 $\times 10^{-4}$ \\
$(2p_02p_0,1s1s,2s^{\prime \prime }2s^{\prime \prime })$ & 5.72 & 5.72 & 2.80
& $\;$0.34295 19455 00106 69550 17114 73595 $\times 10^{-4}$ \\
$I_R$ & & & &         $\;$0.11173 73454 25881 89473 77487 07305 $\times 10^{-4}$ \\
$I_L$ & & & &         $\;$0.23121 46000 74224 80076 39627 66290 $\times 10^{-4}$ \\
$I_{\theta,1}$ & & & &$\;$0.11173 73454 25881 89473 77487 07305 $\times 10^{-4}$ \\
$I_{\theta,2}$ & & & &   -0.11173 73454 25881 89473 77487 07305 $\times 10^{-4}$ \\
$1s3d_0^{\prime \prime },1s1s,1s3d_0^{\prime \prime })$ & 4.26 & 5.72 & 4.26
& $\;$0.79465 21564 41320 36387 60727 96090 $\times 10^{-7}$ \\
$I_R$ & & & &            -0.21880 09534 55753 28525 55400 19381 $\times 10^{-6}$ \\
$I_L$ & & & &            -0.29826 61690 99885 32164 31472 98990 $\times 10^{-6}$ \\
$I_{\theta,1}$ & & & &$\;$0.13058 81740 51792 34673 23175 26295 $\times 10^{-5}$ \\
$I_{\theta,2}$ & & & &   -0.70934 94023 18152 82403 68806 64967 $\times 10^{-6}$ \\
$(3d_03d_0^{\prime \prime },3d_03d_0,3d_03d_0^{\prime \prime })$ & 4.26 &
5.72 & 4.26 &         $\;$0.19195 81649 30384 83368 39585 45964 $\times 10^{-4}$ \\
$I_R$ & & & &         $\;$0.61065 90592 79574 79881 70039 43996 $\times 10^{-4}$ \\
$I_L$ & & & &            -0.42085 49950 53725 04453 18261 50330 $\times 10^{-4}$ \\
$I_{\theta,1}$ & & & &$\;$0.27276 90899 85752 00672 61438 66556 $\times 10^{-5}$ \\
$I_{\theta,2}$ & & & &   -0.25122 80829 40401 21273 83363 43577 $\times 10^{-5}$ \\
$(2p_12p_1^{\prime \prime },1s1s,1s1s)$ & 4.26 & 5.72 & 5.72 & $\;$0.33657
95725 13801 82596 65603 91000 $\times 10^{-5}$ \\
$I_R$ & & & &         $\;$0.63013 65119 15114 04032 65141 14020 $\times 10^{-6}$ \\
$I_L$ & & & &         $\;$0.27356 59213 22290 42193 39089 79598 $\times 10^{-5}$ \\
$I_{\theta,1}$ & & & &   -0.11736 13570 57355 44791 04871 13728 $\times 10^{-6}$ \\
$I_{\theta,2}$ & & & &$\;$0.11736 13570 57355 44791 04871 13728 $\times 10^{-6}$ \\
$(3d_23d_2,3d_13d_1^{\prime \prime },3d_23d_2^{\prime \prime })$ & 5.72 &
4.26 & 4.26 & $\;$0.40399 59434 35959 51311 79010 05972 $\times 10^{-4}$ \\
$I_R$ & & & &             -0.19368 05721 49451 05935 73106 79170 $\times 10^{-3}$ \\
$I_L$ & & & &          $\;$0.23388 47911 23223 44209 78868 70466 $\times 10^{-3}$ \\
$I_{\theta,1}$ & & & &    -0.24807 70502 16847 33560 27030 35045 $\times 10^{-6}$ \\
$I_{\theta,2}$ & & & & $\;$0.44345 24200 40415 90681 66123 36021 $\times 10^{-6}$ \\
$(2p_{-1}2p_{-1}^{\prime \prime },1s1s,1s1s)$ & 4.26 & 5.72 & 5.72 & $\;$%
0.33657 95725 13801 82596 65603 91000 $\times 10^{-5}$ \\
$I_R$ & & & &          $\;$0.63013 65119 15114 04032 65141 14020 $\times 10^{-6}$ \\
$I_L$ & & & &          $\;$0.27356 59213 22290 42193 39089 79598 $\times 10^{-5}$ \\
$I_{\theta,1}$ & & & &    -0.11736 13570 57355 44791 04871 13728 $\times 10^{-6}$ \\
$I_{\theta,2}$ & & & & $\;$0.11736 13570 57355 44791 04871 13728 $\times 10^{-6}$ \\
\hline\hline
\end{tabular}}
\end{table}
\end{center}

\begin{center}
\begin{table}[tbp]
\caption{\footnotesize Continuation of Table I.} 
\scalebox{0.75}{
\begin{tabular}{ccccc}
\hline\hline
Charge distribution & $\omega_1 $ & $\omega_2 $ & $\omega_3$ & $I_{KE}$ \\
\hline
$(3d_{-2}3d_{-2},1s1s,3d_{-2}3d_{-2}^{\prime \prime })$ & 5.72 & 5.72 & 4.26
& $\;$0.87337 99889 07121 70685 75959 02696 $\times 10^{-5}$ \\ 
$I_R$ & & & &           $\;$0.17118 64593 56751 55706 87931 29180 $\times 10^{-5}$ \\
$I_L$ & & & &           $\;$0.69760 53686 99731 53384 13885 67396 $\times 10^{-5}$ \\
$I_{\theta,1}$ & & & &     -0.59358 77640 22225 26505 53260 45002 $\times 10^{-7}$ \\
$I_{\theta,2}$ & & & &  $\;$0.10524 03849 08608 68597 96746 65703 $\times 10^{-6}$ \\
$(2s3p_{1},3p_12s,2s2s)$ & 5.72 & 5.72 & 5.72 & -0.96126 58504 92859 68535 74870 23101 $\times 10^{-8}$ \\
$I_R$ & & & &              -0.26406 66877 11357 08490 91368 00092 $\times 10^{-7}$ \\ 
$I_L$ & & & &              -0.16794 01026 62071 11637 33880 97782 $\times 10^{-7}$ \\ 
$I_{\theta,1}$ & & & &     -0.26818 32640 02419 46684 04183 87723 $\times 10^{-8}$ \\
$I_{\theta,2}$ & & & &  $\;$0.11078 83777 31277 50487 07358 87663 $\times 10^{-7}$ \\
$I_{\varphi} $ & & & &  $\;$0.25191 01539 93106 67456 00821 46673 $\times 10^{-7}$ \\
$(3s3p_{-1},3p_{-1}3s,3s3s)$ & 5.72 & 5.72 & 5.72 & -0.15647 41740 67290 92853 05383 17014 $\times 10^{-7}$ \\
$I_R$ & & & &              -0.33982 21991 19204 52306 98018 90765 $\times 10^{-7}$ \\    
$I_L$ & & & &              -0.18334 80250 51913 59453 92635 73750 $\times 10^{-7}$ \\    
$I_{\theta,1}$ & & & &     -0.30715 98126 09173 86174 12473 94987 $\times 10^{-8}$ \\
$I_{\theta,2}$ & & & &  $\;$0.12238 99937 86874 18344 37565 26374 $\times 10^{-7}$ \\
$I_{\varphi} $ & & & &  $\;$0.27502 20375 77870 39180 88953 60626 $\times 10^{-7}$ \\
$(3d_03d_{1},3d_0^{\prime \prime }3d_{-1}^{\prime \prime },3d_0^{\prime \prime }3d_0^{\prime \prime })$ 
& 5.72 & 2.80 & 2.80 &  $\;$0.36300 14939 30970 01750 58375 17995 $\times 10^{-4}$ \\
$I_R$ & & & &           $\;$0.54502 58949 23235 33885 08159 23174 $\times 10^{-4}$ \\        
$I_L$ & & & &           $\;$0.28430 17983 58119 41570 36818 06466 $\times 10^{-4}$ \\                 
$I_{\theta,1}$ & & & &  $\;$0.19453 49183 43328 87427 09991 55472 $\times 10^{-3}$ \\
$I_{\theta,2}$ & & & &     -0.17963 57860 90172 44749 45254 29761 $\times 10^{-3}$ \\
$I_{\varphi} $ & & & &     -0.61531 75218 81949 00481 33974 68749 $\times 10^{-4}$ \\
$(3d_23d_{1},3d_13d_{2}^{\prime \prime },2p_12p_1^{\prime \prime })$
& 5.72 & 4.26 & 4.26 &     -0.19182 61244 51845 68503 49191 45009 $\times 10^{-6}$ \\
$I_R$ & & & &              -0.19517 83595 75463 95185 57109 41450 $\times 10^{-6}$ \\                         
$I_L$ & & & &              -0.14474 10388 17212 64231 52661 58962 $\times 10^{-6}$ \\                             
$I_{\theta,1}$ & & & &  $\;$0.28724 29747 72837 35917 80423 61548 $\times 10^{-6}$ \\
$I_{\theta,2}$ & & & &     -0.55793 77503 27283 93843 61468 20324 $\times 10^{-7}$ \\
$I_{\varphi} $ & & & &     -0.83355 92579 92780 56198 36972 41143 $\times 10^{-7}$ \\
$(4d_14d_{2},4d_{-1}4d_{-2},2s2s)$
& 5.72 & 5.72 & 5.72 &  $\;$0.14150 98227 95650 24380 54297 78353 $\times 10^{-7}$ \\
$I_R$ & & & &           $\;$0.14564 61757 32755 66307 67419 84589 $\times 10^{-7}$ \\                         
$I_L$ & & & &           $\;$0.99208 16567 62168 55169 02859 42598 $\times 10^{-8}$ \\                             
$I_{\theta,1}$ & & & &  $\;$0.18112 28796 99048 25099 20281 00859 $\times 10^{-7}$ \\
$I_{\theta,2}$ & & & &     -0.17079 27148 25336 90403 74916 44503 $\times 10^{-7}$ \\
$I_{\varphi} $ & & & &     -0.11367 46834 87033 62139 48772 56852 $\times 10^{-7}$ \\
\hline\hline
\end{tabular}}
\end{table} 
\end{center}


\section{Acknowledgments}

The author is indebted to James S. Sims for interesting discussions about
the Kolos and Roothaan transformation and the three-electron kinetic energy integrals. 
The author would like to thank an unknown Reviewer of this paper for the careful reading and 
valuable advice. 

\newpage

\begin{appendix}
\section*{Appendix A: Evaluation of the kinetic energy contribution ${\bf I}_{\bf \theta ,1}\textbf{(1)}$} 

\setcounter{equation}{0}

\renewcommand{\theequation}{A.\arabic{equation}}

Let us define the first angular kinetic energy integral and at the same time expand the 
product of spherical harmonics of electron 3: 

\begin{multline} 
I_{\theta ,1}(1)=\left\langle \phi (\mathbf{r}_1)\phi (\mathbf{r}_2)\phi (\mathbf{r}%
_3)r_{13}|-\frac{r_2}{r_1r_{12}}\frac{\cos {\theta _2}}{\sin {\theta _1}}%
\frac{\partial ^2}{\partial \theta _1\partial r_{12}}|\phi (\mathbf{r}%
_1)\phi (\mathbf{r}_2)\phi (\mathbf{r}_3)r_{12}\right\rangle = \\ 
{}\frac 1{(4\pi )^{1/2}}\sum_{L_3=|l_3-l_3^{\prime }|}^{l_3+l_3^{\prime
}}{}C_{L_3}(2L_3+1)^{1/2}\int_0^\infty r_1^{N_1}e^{-\omega _1r_1}dr_1 
\int_0^\infty r_2^{N_2+2}e^{-\omega_2r_2}r_{12}^{-1}dr_2 \\
\times \int_0^\infty r_3^{N_3+1}e^{-\omega _3r_3}r_{13}dr_3 
\int_0^\pi \int_0^{2\pi }Y_{L_3}^{M_3}(\theta _3,\varphi _3)\sin (\theta_3)d\theta_3d\varphi_3 \\
\times \int_0^\pi \int_0^{2\pi }\cos {\theta _2}Y_{l_2}^{m_2^*}(\theta
_2,\varphi_2)Y_{l_2^{\prime }}^{m_2^{\prime }}(\theta _2,\varphi_2)\sin (\theta_2)d\theta_2d\varphi_2\\
\times \int_0^\pi \int_0^{2\pi }\frac 1{\sin {\theta _1}}Y_{l_1}^{m_1^{*}}(%
\theta _1,\varphi_1)\frac \partial {\partial \theta _1}Y_{l_1^{\prime
}}^{m_1^{\prime }}(\theta _1,\varphi _1)\sin (\theta _1)d\theta _1d\varphi_1 , 
\end{multline}
   
\noindent with $C_{L_3}=C^{L_3}(l_3^{\prime },m_3^{\prime };l_3,m_3)$. The derivative of a spherical harmonic $%
Y_{l_1}^{m_1}(\theta _1,\varphi_1)$ with respect to the polar angle $\theta _1$
\cite[Eq. (5.8.2(1))]{Varshalovich} is: 

\begin{equation}
\frac{\partial Y_{l_1^{\prime }}^{m_1^{\prime }}(\theta _1,\varphi _1)}{%
\partial \theta _1}=f_{1a} Y_{l_1^{\prime }}^{m_1^{\prime
}+1}(\theta _1,\varphi_1)e^{-i\varphi _1} - f_{1b} Y_{l_1^{\prime }}^{m_1^{\prime
}-1}(\theta _1,\varphi_1) e^{i\varphi _1},
\end{equation}

\noindent with
\begin{equation}
f_{1a}=\frac 12[(l_1^{\prime }+m_1^{\prime }+1)(l_1^{\prime }-m_1^{\prime
})]^{1/2},\qquad f_{1b}=\frac 12[(l_1^{\prime }-m_1^{\prime }+1)(l_1^{\prime
}+m_1^{\prime })]^{1/2}. 
\end{equation}
\\
\noindent Multiplying this by the complex conjugate $Y_{l_1}^{m_1*}(\theta _1,\varphi _1)$ and
linearizing the product of spherical harmonics using Eq. (5) we obtain:
\begin{multline}
Y_{l_1}^{m_1*}(\theta _1,\varphi _1)\frac{\partial Y_{l_1^{\prime
}}^{m_1^{\prime }}(\theta _1,\varphi _1)}{\partial \theta _1}=\frac 1{(4\pi
)^{1/2}}\sum_{L_1=|l_1^{\prime }-l_1|}^{l_1^{\prime }+l_1}{}\left[
2L_1+1\right] ^{1/2} \Big\{ f_{1a}C_{L_1a}Y_{L_1}^{M_1+1}(\theta _1,\varphi _1)e^{-i\varphi_1}\\
-f_{1b}C_{L_1b}Y_{L_1}^{M_1-1}(\theta _1,\varphi _1)e^{i\varphi _1} \Big\},  
\end{multline}

\noindent with
\begin{equation}
C_{L_1a}=C^{L_1}(l_1^{\prime },m_1^{\prime }+1;l_1,m_1),\qquad
C_{L_1b}=C^{L_1}(l_1^{\prime },m_1^{\prime }-1;l_1,m_1).
\end{equation}

\noindent Let us use the cosine recursion relation in terms of spherical harmonics \cite{Steinborn}:
\begin{equation}
\cos \theta _2Y_{l_2^{\prime }}^{m_2^{\prime }}(\theta _2,\varphi
_2)=f_{2a}Y_{l_2^{\prime }-1}^{m_2^{\prime }}(\theta _2,\varphi
_2)+f_{2b}Y_{l_2^{\prime }+1}^{m_2^{\prime }}(\theta _2,\varphi _2),
\end{equation}

\noindent for $l_2^{\prime }-1\geq 0$, otherwise the first right-hand-side term vanishes. The factors
are:
\begin{equation}
f_{2a}=\left[ \frac{(l_2^{\prime }+m_2^{\prime })(l_2^{\prime }-m_2^{\prime
})}{(2l_2^{\prime }+1)(2l_2^{\prime }-1)}\right] ^{1/2},\qquad f_{2b}=\left[ 
\frac{(l_2^{\prime }+m_2^{\prime }+1)(l_2^{\prime }-m_2^{\prime }+1)}{%
(2l_2^{\prime }+1)(2l_2^{\prime }+3)}\right] ^{1/2}. 
\end{equation}
\\
\noindent Multiplying again by $Y_{l_2}^{m_2*}(\theta _2,\varphi _2)$ and expanding the
products of spherical harmonics using Eq. (5) we obtain:
\begin{multline}
Y_{l_2}^{m_2*}(\theta _2,\varphi _2)\cos \theta _2Y_{l_2^{\prime
}}^{m_2^{\prime }}(\theta _2,\varphi _2)= 
\frac 1{(4\pi )^{1/2}}\sum_{L_2=|l_2^{\prime }-1-l_2|}^{l_2^{\prime
}-1+l_2}{}\left[ 2L_2+1\right] ^{1/2}f_{2a}C_{L_2a}Y_{L_2}^{M_2}(\theta
_2,\varphi _2) \\
+\frac 1{(4\pi )^{1/2}}\sum_{L_2^{\prime }=|l_2^{\prime
}+1-l_2|}^{l_2^{\prime }+1+l_2}{}\left[ 2L_2^{\prime }+1\right]
^{1/2}f_{2b}C_{L_2b}Y_{L_2^{\prime }}^{M_2^{\prime }}(\theta _2,\varphi _2), 
\end{multline}
with
\begin{equation}
C_{L_2a}=C^{L_2}(l_2^{\prime }-1,m_2^{\prime };l_2,m_2),\qquad
C_{L_2b}=C^{L_2^{\prime }}(l_2^{\prime }+1,m_2^{\prime };l_2,m_2). 
\end{equation}
\\
Altogether 
\begin{multline}
-\frac 1{\sin {\theta _1}} Y_{l_3}^{m_3*}(\theta _3,\varphi
_3)Y_{l_3^{\prime }}^{m_3^{\prime }}(\theta _3,\varphi _3) 
Y_{l_2}^{m_2*}(\theta _2,\varphi _2)\cos \theta _2Y_{l_2^{\prime
}}^{m_2^{\prime }}(\theta _2,\varphi _2) 
Y_{l_1}^{m_1*}(\theta _1,\varphi _1)\frac{\partial Y_{l_1^{\prime
}}^{m_1^{\prime }}(\theta _1,\varphi _1)}{\partial \theta _1} = \\
\frac{1}{(4\pi)^{3/2}} 
\sum_{L_1=|l_1^{\prime }-l_1|}^{l_1^{\prime }+l_1}{}\sum_{L_3=|l_3^{\prime
}-l_3|}^{l_3^{\prime }+l_3}\left[ (2L_1+1)(2L_3+1)\right] ^{1/2} 
C_{L_3}Y_{L_3}^{M_3}(\theta _3,\varphi _3) \left( -\frac 1{\sin \theta_1}\right) \\
\times \left\{ \sum_{L_2=|l_2^{\prime }-1-l_2|}^{l_2^{\prime
}-1+l_2}{}\left[ 2L_2+1\right] ^{1/2} f_{2a} C_{L_2a} \Big( f_{1a}C_{L_1a}Y_{L_1}^{M_1+1}(\theta _1,\varphi
_1)e^{-i\varphi_1}Y_{L_2}^{M_2}(\theta _2,\varphi _2) \right.  \\
 -f_{1b}C_{L_1b}Y_{L_1}^{M_1-1}(\theta _1,\varphi
_1)e^{i\varphi_1}Y_{L_2}^{M_2}(\theta _2,\varphi _2)\Big) \\
+\sum_{L_2^{\prime }=|l_2^{\prime }+1-l_2|}^{l_2^{\prime }+1+l_2}{}\left[
2L_2^{\prime }+1\right] ^{1/2} f_{2b} C_{L_2b} \left(
f_{1a}C_{L_1a}Y_{L_1}^{M_1+1}(\theta _1,\varphi _1)e^{-i\varphi_1}Y_{L_2^{\prime
}}^{M_2}(\theta _2,\varphi _2)\right. \\
\left. -f_{1b}C_{L_1b}Y_{L_1}^{M_1-1}(\theta _1,\varphi
_1)e^{i\varphi_1}Y_{L_2^{\prime }}^{M_2}(\theta _2,\varphi _2)\right) \Bigg\}. 
\end{multline}

Passing the z-axis through the $r_1$ coordinate, this rotation transforms the
variables $\theta _2\rightarrow \theta _{12}$, and $\phi
_2\rightarrow \phi _{12}$ . This may be understood graphically in Figure 1.
This rotation of the z-axis produces a rotation in the spherical harmonic $Y_{L_2}^{M_2}(\theta _2,\varphi _2)$ 
(see the details in the previous articles of this series \cite[Eq. (22)]{Ruiz3e}):
\begin{equation}
Y_{L_{2}}^{M_{2}}(\theta _{2},\varphi _{2})=\left( \frac{4\pi}{2L_2+1}\right)^{1/2}
\sum_{M_{2}^{\prime}=-L_{2}}^{L_{2}}Y_{L_{2}}^{M_{2}}(\theta _{1},\varphi
_{1})Y_{l_{2}}^{M_{2}^{\prime }}({\theta_{12},\varphi_{12}}) . 
\end{equation}

Afterwards, 
the integration over $\varphi _{12}$ leads to $4\pi  Y_{L_2}^{M_2}(\theta _1,\varphi
_1)$ and to Legendre polynomials of the form $P_{L_2}(\cos\theta_{12})$:

\begin{multline}
I_{\theta ,1}(1)=\frac{1}{(4\pi)^{1/2}}\sum_{L_1=|l_1^{\prime }-l_1|}^{l_1^{\prime }+l_1}{}
\sum_{L_3=|l_3^{\prime}-l_3|}^{l_3^{\prime }+l_3}\Big[ (2L_1+1)(2L_3+1)\Big] ^{1/2} 
C_{L_3} \left( -\frac 1{\sin \theta_1}\right) \\ 
\times \int_0^\infty r_1^{N_1}e^{-\omega _1r_1}dr_1 \int_0^\infty r_2^{N_2+2}e^{-\omega
_2r_2}dr_2\int_0^\infty r_3^{N_3+1}e^{-\omega _3r_3}dr_3 
\int_0^\pi \int_0^{2\pi }
Y_{L_3}^{M_3}(\theta _3,\varphi _3)\sin\theta _3d\theta _3d\varphi_3 \\
\times \left\{ \sum_{L_2=|l_2^{\prime }-1-l_2|}^{l_2^{\prime
}-1+l_2}{}\left[ 2L_2+1\right] ^{1/2} f_{2a} C_{L_2a}  
\int_0^{\pi} \frac 12 r_{12}^{-1}P_{L_2}(\cos\theta_{12}) \sin\theta_{12}d\theta_{12} \right. \\
\times \Big( f_{1a}C_{L_1a}\int_0^\pi \int_0^{2\pi }
Y_{L_1}^{M_1+1}(\theta _1,\varphi_1)e^{-i\varphi_1}Y_{L_2}^{M_2}(\theta _1,\varphi _1)
\sin\theta _1d\theta _1d\varphi_1   \\
 -f_{1b}C_{L_1b}\int_0^\pi \int_0^{2\pi }Y_{L_1}^{M_1-1}(\theta _1,\varphi
_1)e^{i\varphi_1}Y_{L_2}^{M_2}(\theta _1,\varphi _1)\sin\theta _1d\theta _1d\varphi_1 \Big) \\
+\sum_{L_2^{\prime }=|l_2^{\prime }+1-l_2|}^{l_2^{\prime }+1+l_2}{}\left[
2L_2^{\prime }+1\right] ^{1/2} f_{2b} C_{L_2b}
\int_0^{\pi} \frac 12 r_{12}^{-1}P_{L_2^{\prime}}(\cos\theta_{12}) \sin\theta_{12}d\theta_{12} \\
\times \left( f_{1a}C_{L_1a}
\int_0^\pi \int_0^{2\pi }Y_{L_1}^{M_1+1}(\theta _1,\varphi _1)e^{-i\varphi_1}Y_{L_2^{\prime
}}^{M_2}(\theta _1,\varphi _1)\sin\theta_1d\theta _1d\varphi_1 \right. \\
\left.  -f_{1b}C_{L_1b}
\int_0^\pi \int_0^{2\pi }Y_{L_1}^{M_1-1}(\theta _1,\varphi
_1)e^{i\varphi_1}Y_{L_2^{\prime }}^{M_2}(\theta _1,\varphi _1)\sin\theta_1 d\theta _1d\varphi_1
 \right) \Bigg\}. 
\end{multline}

Now the spherical harmonics with the same arguments can be expanded again,
see Eq. (5).
Note we need the complex conjugates of $Y_{L_2}^{M_2}$ and $Y_{L_2^{\prime}}^{M_2}$, see Eq. (4). 
In addition we rotate $Y_{L_3}^{M_3}$ according to Eq. (A.11) and get another factor $4\pi$:

\begin{multline}
I_{\theta ,1}(1)=(-1)^{M_2}\sum_{L_1=|l_1^{\prime }-l_1|}^{l_1^{\prime
}+l_1}\sum_{L_3=|l_3^{\prime }-l_3|}^{l_3^{\prime }+l_3}{}\Big[
(2L_1+1)(2L_3+1)\Big] ^{1/2}C_{L_3}\left( -\frac 1{\sin \theta _1}\right) \\
\times \Bigg\{ \sum_{L_2=|l_2^{\prime }-1-l_2|}^{l_2^{\prime
}-1+l_2}{}\sum_{K_1=|L_2-L_1|}^{L_1+L_2}
\Big[ (2L_2+1)(2K_1+1) \Big]^{1/2} f_{2a} C_{L_2a}\\
\times \int_0^\infty r_1^{N_1}e^{-\omega _1r_1}dr_1 
\int_0^\infty r_2^{N_2+2}e^{-\omega
_2r_2}dr_2\int_0^\infty r_3^{N_3+1}e^{-\omega _3r_3}dr_3 \\
\times \int_0^{\pi} \frac 12 r_{12}^{-1}P_{L_2}(\cos\theta_{12}) \sin\theta_{12}d\theta_{12}  
\int_0^{\pi} \frac 12 r_{13} P_{L_3}(\cos\theta_{13}) \sin\theta_{13}d\theta_{13} \\ 
\times \Big( f_{1a} C_{L_1a}C_{K_1a}
\int_0^\pi \int_0^{2\pi} Y_{L_3}^{M_3}(\theta _1,\varphi_1)Y_{K_1}^{M_1+M_2+1}(\theta _1,\varphi _1) 
e^{-i\varphi_1} \sin\theta_1d\theta_1d\varphi_1 \\
 -f_{1b}C_{L_1b}C_{K_1b}\int_0^\pi \int_0^{2\pi} Y_{L_3}^{M_3}(\theta _1,\varphi_1)
Y_{K_1}^{M_1+M_2-1}(\theta _1,\varphi _1)e^{i\varphi_1} \sin\theta_1d\theta_1d\varphi_1 \Big)  \\
+\sum_{L_2^{\prime }=|l_2^{\prime }+1-l_2|}^{l_2^{\prime
}+1+l_2}{}\sum_{K_1^{\prime }=|L_2^{\prime }-L_1|}^{L_1+L_2^{\prime }}
\Big[(2L_2^{\prime }+1)(2K_1^{\prime }+1)\Big]^{1/2} f_{2b} C_{L_2b}\\
\times \int_0^\infty r_1^{N_1}e^{-\omega _1r_1}dr_1 
\int_0^\infty r_2^{N_2+2}e^{-\omega
_2r_2}dr_2\int_0^\infty r_3^{N_3+1}e^{-\omega _3r_3}dr_3 \\
\times \int_0^{\pi} \frac 12 r_{12}^{-1}P_{L_2^{\prime}}(\cos\theta_{12}) \sin\theta_{12}d\theta_{12}  
\int_0^{\pi} \frac 12 r_{13} P_{L_3}(\cos\theta_{13}) \sin\theta_{13}d\theta_{13} \\ 
\times \Big( f_{1a}C_{L_1a}C_{K_1c}\int_0^\pi \int_0^{2\pi} Y_{L_3}^{M_3}(\theta _1,\varphi_1)
Y_{K_1^{\prime }}^{M_1+M_2+1}(\theta _1,\varphi _1) e^{-i\varphi_1} \sin\theta_1d\theta_1d\varphi_1 \\
  -f_{1b}C_{L_1b}C_{K_1d}\int_0^\pi \int_0^{2\pi} Y_{L_3}^{M_3}(\theta _1,\varphi_1)
Y_{K_1^{\prime }}^{M_1+M_2-1}(\theta _1,\varphi _1) e^{i\varphi_1}\sin\theta_1d\theta_1d\varphi_1 
\Big) \Bigg\} , 
\end{multline}

and
\begin{eqnarray}
C_{K_1a} &=&C^{K_1}(L_1,M_1+1;L_2,-M_2),\qquad C_{K_1b}=C^{K_1}(L_1,M_1-1;L_2,-M_2),  
\nonumber \\
C_{K_1c} &=&C^{K_1^{\prime }}(L_1,M_1+1;L_2^{\prime },-M_2),\qquad
C_{K_1d}=C^{K_1^{\prime }}(L_1,M_1-1;L_2^{\prime },-M_2). 
\end{eqnarray}
We write the radial part in terms of radial three-electron integrals $J$ and linearize again the 
products of spherical harmonics with same arguments.  
Note that the radial three-electron integral fulfill the symmetry relation:
\begin{multline}
J(N_1-1,N_2+1,N_3;\omega _1,\omega _2,\omega _3;-1,1;L_2,L_3)= 
J(N_1-1,N_3,N_2+1;\omega _1,\omega _3,\omega _2;1,-1;L_3,L_2), 
\end{multline}

\noindent obtaining: 
\begin{multline}
I_{\theta ,1}(1)= \frac{(-1)^{M_2+M_3}}{(4\pi )^{1/2}}\sum_{L_1=|l_1^{\prime }-l_1|}^{l_1^{\prime
}+l_1}{}\sum_{L_3=|l_3^{\prime }-l_3|}^{l_3^{\prime }+l_3}\Big[
(2L_1+1)(2L_3+1)\Big] ^{1/2}C_{L_3}\left( -\frac 1{\sin \theta _1}\right) \\
\times \left\{ \sum_{L_2=|l_2^{\prime }-1-l_2|}^{l_2^{\prime
}-1+l_2}{}\sum_{K_1=|L_2-L_1|}^{L_1+L_2}\sum_{K_2=|K_1-L_3|}^{K_1+L_3}\right.
\Big[(2L_2+1)(2K_1+1)(2K_2+1)\Big]^{1/2} f_{2a} C_{L_2a}\\
\times \Big( f_{1a}C_{L_1a}C_{K_1a}C_{K_2a}
\int_0^\pi \int_0^{2\pi}Y_{K_2}^{M_1+M_2+M_3+1}(\theta_1,\varphi _1) e^{-i\varphi_1} 
\sin\theta_1d\theta_1d\varphi_1\\
 -f_{1b}C_{L_1b}C_{K_1b}C_{K_2b}\int_0^\pi \int_0^{2\pi}
Y_{K_2}^{M_1+M_2+M_3-1}(\theta_1,\varphi _1)e^{i\varphi_1} 
\sin\theta_1d\theta_1d\varphi_1 \Big) \\
\times J(N_1-1,N_3,N_2+1;\omega _1,\omega _3,\omega _2;1,-1;L_3,L_2) \\
+\sum_{L_2^{\prime }=|l_2^{\prime }+1-l_2|}^{l_2^{\prime
}+1+l_2}{}\sum_{K_1^{\prime }=|L_2^{\prime }-L_1|}^{L_1+L_2^{\prime}}\sum_{K_2^{\prime
}=|K_1^{\prime }-L_3|}^{K_1^{\prime }+L_3}\Big[(2L_2^{\prime
}+1)(2K_1^{\prime }+1)(2K_2^{\prime }+1)\Big]^{1/2} f_{2b} C_{L_2b} \\
\times \Big( f_{1a}C_{L_1a}C_{K_1c}C_{K_2c} \int_0^\pi \int_0^{2\pi}
Y_{K_2^{\prime}}^{M_1+M_2+M_3+1}(\theta _1,\varphi _1)e^{-i\varphi_1} 
\sin\theta_1d\theta_1d\varphi_1 \\
  -f_{1b}C_{L_1b}C_{K_1d}C_{K_2d} \int_0^\pi \int_0^{2\pi} 
Y_{K_2^{\prime}}^{M_1+M_2+M_3-1}(\theta _1,\varphi _1)e^{i\varphi_1} 
\sin\theta_1d\theta_1d\varphi_1 \Big) \\
 \times J(N_1-1,N_3,N_2+1;\omega _1,\omega _3,\omega
_2;1,-1;L_3,L_2^{\prime }) \Bigg\} , 
\end{multline}

\noindent with 
\begin{eqnarray}
C_{K_2a} &=&C^{K_2}(K_1,M_1+M_2+1;L_3,-M_3),\qquad
C_{K_2b}=C^{K_2}(K_1,M_1+M_2-1;L_3,-M_3),   \nonumber \\
C_{K_2c} &=&C^{K_2^{\prime }}(K_1^{\prime },M_1+M_2+1;L_3,-M_3),\qquad
C_{K_2d}=C^{K_2^{\prime }}(K_1^{\prime },M_1+M_2-1;L_3,-M_3) .  \nonumber \\
\end{eqnarray}

Finally, the recursion relation containing the sine function  
Eq. (1) can be applied. This relation is valid for $M\neq 0$. This condition is fulfilled in our cases 
since the above integrals over spherical harmonics include the condition $M_1+M_2+M_3=0$. Therefore 
$M=M_1+M_2+M_3+1$ and $M^{\prime }=M_1+M_2+M_3-1$, what concretely means $M^{\prime}\neq 0$ 
and $M\neq 0$. Multiplying now by $\big(1/\sin\theta _1\big)$ and using Eq. (A.16) we have:

\begin{multline}
I_{\theta ,1}(1)=\frac{(-1)^{M_2+M_3}}{(4\pi )^{1/2}}\sum_{L_1=|l_1^{\prime
}-l_1|}^{l_1^{\prime }+l_1}{}\sum_{L_3=|l_3^{\prime }-l_3|}^{l_3^{\prime
}+l_3}\Big[ (2L_1+1)(2L_3+1)\Big] ^{1/2}C_{L_3} \\
\times \left\{ \sum_{L_2=|l_2^{\prime }-1-l_2|}^{l_2^{\prime
}-1+l_2}{}\sum_{K_1=|L_2-L_1|}^{L_1+L_2}%
\sum_{K_2=|K_1-L_3|}^{K_1+L_3} \Bigg[ \frac{(2K_2+1)}{(2K_2+3)}\Bigg]^{1/2} 
\Big[(2L_2+1)(2K_1+1)(2K_2+1)\Big]^{1/2}\right. \\
\times f_{2a} C_{L_2a} \left( f_{1a}f_b(K_2,M)C_{L_1a}C_{K_1a}C_{K_2a}\int_0^\pi \int_0^{2\pi
}Y_{K_2+1}^{M-1}(\theta _1,\varphi _1) \sin\theta _1 d\theta _1d\varphi_1  \right. \\
+f_{1a}f_a(K_2,M)C_{L_1a}C_{K_1a}C_{K_2a}\int_0^\pi \int_0^{2\pi
}Y_{K_2+1}^{M+1}(\theta _1,\varphi _1)e^{-2i\varphi _1}\sin\theta_1 d\theta _1d\varphi_1 \\
-f_{1b}f_b(K_2,M^{\prime})C_{L_1b}C_{K_1b}C_{K_2b}\int_0^\pi \int_0^{2\pi
}Y_{K_2+1}^{M^{\prime}-1}(\theta _1,\varphi _1)e^{2i\varphi_1 }\sin\theta_1 d\theta _1d\varphi_1 \\
\left. -f_{1b}f_a(K_2,M^{\prime})C_{L_1b}C_{K_1b}C_{K_2b}\int_0^\pi \int_0^{2\pi
}Y_{K_2+1}^{M^{\prime}+1}(\theta _1,\varphi _1)\sin\theta_1 d\theta _1d\varphi_1 \right) \\
\times J(N_1-1,N_3,N_2+1;\omega _1,\omega _3,\omega _2;1,-1;L_3,L_2) \\
+\sum_{L_2^{\prime }=|l_2^{\prime }+1-l_2|}^{l_2^{\prime
}+1+l_2}{}\sum_{K_1^{\prime }=|L_2^{\prime }-L_1|}^{L_1+L_2^{\prime}}
\sum_{K_2^{\prime}=|K_1^{\prime}-L_3|}^{K_1^{\prime}+L_3}
\Bigg[ \frac{(2K_2^{\prime}+1)}{(2K_2^{\prime}+3)}\Bigg]^{1/2}
\Big[ (2L_2+1)(2K_1^{\prime }+1)2K_2^{\prime}+1)\Big]^{1/2} \\
\times f_{2b} C_{L_2b} \left( f_{1a}f_b(K_2^{\prime},M)C_{L_1a}C_{K_1c}C_{K_2c}\int_0^\pi \int_0^{2\pi
}Y_{K_2^{\prime }+1}^{M-1}(\theta _1,\varphi _1)\sin\theta_1d\theta _1d\varphi_1\right. \\
+f_{1a}f_a(K_2^{\prime},M)C_{L_1a}C_{K_1c}C_{K_2c}\int_0^\pi \int_0^{2\pi
}Y_{K_2^{\prime }+1}^{M+1}(\theta _1,\varphi _1)e^{-2i\varphi _1}\sin\theta_1 d\theta _1d\varphi_1 \\
-f_{1b}f_b(K_2^{\prime},M^{\prime})C_{L_1b}C_{K_1d}C_{K_2d}\int_0^\pi \int_0^{2\pi
}Y_{K_2^{\prime }+1}^{M^{\prime}-1}(\theta _1,\varphi _1)e^{2i\varphi _1}\sin\theta_1 d\theta _1d\varphi_1 \\
\left. -f_{1b}f_a(K_2^{\prime},M^{\prime})C_{L_1b}C_{K_1d}C_{K_2d}\int_0^\pi \int_0^{2\pi
}Y_{K_2^{\prime }+1}^{M^{\prime}+1}(\theta _1,\varphi _1) \sin\theta_1d\theta_1d\varphi_1 \right) \\
\times J(N_1-1,N_3,N_2+1;\omega _1,\omega _3,\omega_2;1,-1;L_3,L_2^{\prime }) \Bigg\} , 
\end{multline}

\noindent with 
\begin{eqnarray}
f_a(K,M)&=&\frac 1{2M}[(K+M+2)(K+M+1)]^{1/2}, \nonumber \\
f_b(K,M)&=&\frac 1{2M}[(K-M+2)(K-M+1)]^{1/2},
\end{eqnarray}

\noindent and 
$M=M_1+M_2+M_3+1$ and $M^{\prime }=M_1+M_2+M_3-1.$

The first kind of integrals over spherical harmonics are:
\begin{equation}
\int_0^\pi \int_0^{2\pi }Y_{K_1+1}^{M_1+M_2+M_3}(\theta _1,\phi _1)d\Omega
_1=\delta (K_1+1,0)\delta (M_1+M_2+M_3,0)=0
\end{equation}
and the same holds for $Y_{K_1^{\prime}+1}^{M_1+M_2+M_3}(\theta _1,\phi _1)$. 
These integrals are zero because $K_1+1\geq 0$. The following integrals are
evaluated integrating over the associated Legendre functions:
\begin{multline}
B_{1}(L,M+n)=\int_0^\pi \int_0^{2\pi }Y_L^{M+n}(\theta _1,\phi
_1)e^{-ni\phi _1}d\Omega _1 \\
=\pi ^{1/2}(2L+1)^{1/2}\left[ \frac{(L-M-n)!}{(L+M+n)!}\right]
^{1/2}\int_0^\pi P_L^{M+n}(\cos \theta _1)\sin \theta _1d\theta _1 , 
\end{multline}

\noindent using the algorithm developed by Wong \cite{Wong} for the overlap
integral over associated Legendre functions, which can be effectively reduced to: 
\begin{equation}
\int_0^\pi P_{L}^{M}(\cos \theta )\sin
(\theta )d\theta =\sum_{p=0}^{p_{\max }} a_{L,M}^{p} 
\frac{\Gamma \left( \frac 12(L-M-2p+1)\right)
\Gamma \left( \frac 12(M+2p+2)\right) }{\Gamma \left( \frac
12(L+3)\right) }
\end{equation}

\noindent with the coefficients:

\begin{equation}
a_{L_{,}M}^p=\frac{(-1)^p(L+M)!}{2^{M+2p}(M+p)!p!(L-M-2p)!}.
\end{equation}

\bigskip
\noindent $\Gamma $ are Gamma functions. $p_{\max }=[(L-M)/2]$ is the integral part of 
$(L-M)/2$. This formula is valid for $0\leq M_1\leq L_1$. For $M\geq L$ 
the integral is zero, see Ref. \cite
{Wong}. If $M$ is negative the following formula is used:

\begin{equation}
P_L^{-M}(\cos \theta _1)=(-1)^M\frac{(L-M)!}{(L+M)!}P_L^M(\cos \theta _1).
\end{equation}

Finally the programmable expression is:
\begin{multline}
I_{\theta ,1}(1)=\frac{(-1)^{M_2+M_3}}{(4\pi )^{1/2}} \delta (M_1+M_2+M_3,0) 
\sum_{L_1=|l_1^{\prime }-l_1|}^{l_1^{\prime
}+l_1}{}\sum_{L_3=|l_3^{\prime }-l_3|}^{l_3^{\prime }+l_3}C_{L_3}\left\{
\sum_{L_2=|l_2^{\prime }-1-l_2|}^{l_2^{\prime
}-1+l_2}{}\sum_{K_1=|L_2-L_1|}^{L_1+L_2}\sum_{K_2=|K_1-L_3|}^{K_1+L_3} \right. \\
\times \Bigg[\frac{(2K_2+1)}{(2K_2+3)}\Bigg]^{1/2} 
\Big[(2L_1+1)(2L_2+1)(2L_3+1)(2K_1+1)(2K_2+1)\Big]^{1/2} f_{2a} C_{L_2a}\\
\times \Big(
f_{1a}f_{a}(K_2,M)C_{L_1a}C_{K_1a}C_{K_2a}\;B_{1}(K_2+1,M+1) \\ 
-f_{1b}f_{b}(K_2,M^{\prime})C_{L_1b}C_{K_1b}C_{K_2b}\;B_{1}(K_2+1,M^{\prime
}-1)\Big)  \\
\times J(N_1-1,N_3,N_2+1;\omega _1,\omega _3,\omega _2;1,-1;L_3,L_2) \\
+\sum_{L_2^{\prime }=|l_2^{\prime }+1-l_2|}^{l_2^{\prime
}+1+l_2}{}\sum_{K_1^{\prime }=|L_2^{\prime }-L_1|}^{L_1+L_2^{\prime}}\sum_{K_2^{\prime
}=|K_1^{\prime }-L_3|}^{K_1^{\prime }+L_3}\Bigg[\frac{(2K_2^{\prime}+1)}{(2K_2^{\prime}+3)}\Bigg]^{1/2}\\
\times \Big[ (2L_1+1)(2L_2^{\prime }+1)(2L_3+1)(2K_1^{\prime }+1)(2K_2^{\prime}+1)\Big] ^{1/2} f_{2b} C_{L_2b}\\
\times \Big( f_{1a}f_{a}(K_2^{\prime},M)C_{L_1a}C_{K_1c}C_{K_2c}\;B_{1}(K_2^{\prime }+1,M+1) \\ 
-f_{1b}f_{b}(K_2^{\prime},M^{\prime})C_{L_1b}C_{K_1d}C_{K_2d}\;B_{1}(K_2^{\prime
}+1,M^{\prime }-1)\Big)  \\
 \times J(N_1-1,N_3,N_2+1;\omega _1,\omega _3,\omega_2;1,-1;L_3,L_2^{\prime })\Bigg\} . 
\end{multline}

\end{appendix}

\newpage

\begin{appendix}
\section*{Appendix B: Evaluation of the kinetic energy contribution ${\bf I}_{\bf \theta ,2}\textbf{(1)}$}

\setcounter{equation}{0}

\renewcommand{\theequation}{B.\arabic{equation}}

Let us evaluate the integral over electron 1 of a matrix element 
generated by the operator $\hat T_{\theta ,2}(1)$ and 
configurations including the interelectronic distances $r_{12}$ on the right-hand side and $r_{13}$ on the 
left-hand side: 

\begin{multline}
I_{\theta ,2}(1)=\left\langle \phi (\mathbf{r}_1)\phi (\mathbf{r}%
_2)\phi (\mathbf{r}_3)r_{13}|-\frac 12 \frac{(r_{12}^2-r_1^2-r_2^2)}{r_1^2r_{12}}\cot {%
\theta _1}\frac{\partial ^2}{\partial \theta _1\partial r_{12}}|\phi^{\prime } (%
\mathbf{r}_1)\phi^{\prime } (\mathbf{r}_2)\phi^{\prime } (\mathbf{r}_3)r_{12}\right\rangle .  
\end{multline}

For the evaluation of $I_{\theta ,2}$, we repeat the same steps of Appendix A, Eqs. (A.2-A.5). 
Afterwards, the products of 
spherical harmonics of electrons 2 and 3 should be linearized. Then it follows the rotations of the 
functions of electrons 2 and 3 according Eq. (A.11), and the use of the complex conjugate of $Y_{L_1}^{M_1-1}(%
\theta _1,\varphi _1)$, in order to expand the products of $Y_{L_1}^{-M_1+1*}(\theta
_1,\varphi_1)Y_{L_2}^{M_2}(\theta _1,\varphi_1)$. After all these steps we obtain:
\begin{multline}
I_{\theta ,2}(1)=\frac{(-1)^{M_2+M_3}}{4\sqrt{\pi}}\sum_{L_1=|l_1^{\prime }-l_1|}^{l_1^{\prime
}+l_1}{}\sum_{L_2=|l_2^{\prime }-l_2|}^{l_2^{\prime
}+l_2}{}\sum_{L_3=|l_3^{\prime }-l_3|}^{l_3^{\prime }+l_3} \\
\times \sum_{L=|L_1-L_2|}^{L_1+L_2}{}\sum_{L^{\prime }=|L-L_3|}^{L+L_3}\Big[
(2L_1+1)(2L_2+1)(2L_3+1)(2L+1)(2L^{\prime }+1)\Big] ^{1/2}C_{L_2}C_{L_3} \\
\times \Bigg\{ f_{1a}C_{L_1a}C_{La}C_{Lc}Y_{L^{\prime }}^{M_1+M_2+M_3+1}(\theta
_1,\varphi _1)e^{-i\varphi _1} 
 -f_{1b}C_{L_1b}C_{Lb}C_{Ld}Y_{L^{\prime }}^{M_1+M_2+M_3-1}(\theta _1,\varphi
_1)e^{i\varphi _1}\Bigg\} \\
\times \Big[ J(N_1-2,N_2,N_3;\omega _1,\omega _2,\omega_3;1,1;L_2,L_3) 
-J(N_1,N_2;N_3;\omega _1,\omega _2,\omega _3;-1,1;L_2,L_3) \\
-J(N_1-2,N_2+2,N_3;\omega _1,\omega _2,\omega _3;-1,1;L_2,L_3)\Big], 
\end{multline}

\noindent with
\begin{equation}
f_{1a}=\frac 12[(l_1^{\prime }+m_1^{\prime }+1)(l_1^{\prime }-m_1^{\prime
})]^{1/2},\qquad f_{1b}=\frac 12[(l_1^{\prime }-m_1^{\prime }+1)(l_1^{\prime
}+m_1^{\prime })]^{1/2},  
\end{equation}
and 
\begin{eqnarray}
C_{L_1a}=C^{L_1}(l_1^{\prime },m_1^{\prime }+1;l_1,m_1), &\qquad &
C_{L_1b}=C^{L_1}(l_1^{\prime },m_1^{\prime }-1;l_1,m_1), \nonumber \\ 
C_{L_2}=C^{L_2}(l_2^{\prime },m_2^{\prime };l_2,m_2), & \qquad & 
C_{L_3}=C^{L_3}(l_3^{\prime },m_3^{\prime };l_3,m_3),  \nonumber \\
C_{L_a} =C^L(L_1,M_1+1;L_2,-M_2),& \qquad & C_{L_b}=C^L(L_1,M_1-1;L_2,-M_2), \nonumber \\
C_{L_c} =C^{L^{\prime }}(L_1,M_1+1;L_3,-M_3),& \qquad & C_{L_d}=C^{L^{\prime
}}(L_1,M_1-1;L_3,-M_3). 
\end{eqnarray}

Now let us apply the recursion relation between spherical harmonics with general arguments $L$, $M$ 
including the $\cot \theta $ function 
\cite[Eq. (5.7.)]{Varshalovich}:
\begin{eqnarray}
-\cot \theta Y_{L}^{M}(\theta ,\varphi ) &=&\frac1{2M}
\Big[(L+M+1)(L-M)\Big]^{1/2} Y_L^{M+1}(\theta ,\varphi ) e^{-i\varphi } \nonumber \\
&&+\frac 1{2M}\Big[(L-M+1)(L+M)\Big]^{1/2} Y_L^{M-1}(\theta ,\varphi )e^{i\varphi }, 
\end{eqnarray}
and perform radial integration, which leads to:
\begin{multline}
I_{\theta ,2}(1)=\frac{(-1)^{M_2+M_3}}{4\sqrt{\pi}}\sum_{L_1=|l_1^{\prime
}-l_1|}^{l_1^{\prime }+l_1}{}{}\sum_{L_2=|l_2^{\prime }-l_2|}^{l_2^{\prime
}+l_2}{}\sum_{L_3=|l_3^{\prime }-l_3|}^{l_3^{\prime
}+l_3}\sum_{L=|L_1-L_2|}^{L_1+L_2}\sum_{L^{\prime }=|L-L_3|}^{L+L_3} \\
\times \Big[ (2L_1+1)(2L_2+1)(2L_3+1)(2L+1)(2L^{\prime }+1)\Big]^{1/2}C_{L_2}C_{L_3} \\
\times \Bigg\{ f_{1a}f_{2a}C_{1a}C_{La}C_{Lc}\int_0^\pi \int_0^{2\pi
}Y_{L^{\prime }}^{M_1+M_2+M_3+2}(\theta _1,\varphi _1)e^{-2i\varphi _1}
\sin(\theta_1)d\theta_1d\varphi_1  \\
+f_{1a}f_{2b}C_{1a}C_{La}C_{Lc}\int_0^\pi \int_0^{2\pi }Y_{L^{\prime
}}^{M_1+M_2+M_3}(\theta _1,\varphi _1)\sin(\theta_1)d\theta_1d\varphi_1 \\
-f_{1b}f_{2c}C_{1b}C_{Lb}C_{Ld}\int_0^\pi \int_0^{2\pi }Y_{L^{\prime
}}^{M_1+M_2+M_3}(\theta _1,\varphi _1)\sin(\theta_1)d\theta_1d\varphi_1 \\
-f_{1b}f_{2d}C_{1b}C_{Lb}C_{Ld}\int_0^\pi \int_0^{2\pi }Y_{L^{\prime
}}^{M_1+M_2+M_3-2}(\theta _1,\varphi _1)e^{2i\varphi _1}\sin(\theta_1)d\theta_1d\varphi_1\Bigg\} \\
\times \Big[ J(N_1-2,N_2,N_3;\omega _1,\omega _2,\omega_3;1,1;L_2,L_3) 
-J(N_1,N_2;N_3;\omega _1,\omega _2,\omega _3;-1,1;L_2,L_3) \\
-J(N_1-2,N_2+2,N_3;\omega _1,\omega _2,\omega _3;-1,1;L_2,L_3)\Big]
,
\end{multline}
with $M_1+M_2+M_3=0$, $M=M_1+M_2+M_3+1$ and $M^{\prime }=M_1+M_2+M_3-1$. The factors are: 

\begin{eqnarray}
f_{2a}=\frac 1{2M}[(L^{\prime }+M+1)(L^{\prime }-M)]^{1/2}, & \qquad &  
f_{2b}=\frac 1{2M}[(L^{\prime }-M+1)(L^{\prime }+M)]^{1/2}, \nonumber \\
f_{2c}=\frac 1{2M^{\prime }}[(L^{\prime }+M^{\prime }+1)(L^{\prime
}-M^{\prime })]^{1/2}, & \qquad & 
f_{2d}=\frac 1{2M^{\prime }}[(L^{\prime }+M^{\prime }+1)(L^{\prime
}+M^{\prime })]^{1/2}. \nonumber \\
\end{eqnarray}

The final programmable expression is:
\begin{multline}
I_{\theta ,2}(1)=\frac{(-1)^{M_2+M_3}}{4\sqrt{\pi}}\delta (M_1+M_2+M_3,0) \sum_{L_1=|l_1^{\prime
}-l_1|}^{l_1^{\prime }+l_1}{}{}\sum_{L_2=|l_2^{\prime }-l_2|}^{l_2^{\prime
}+l_2}{}\sum_{L_3=|l_3^{\prime }-l_3|}^{l_3^{\prime
}+l_3}\sum_{L=|L_1-L_2|}^{L_1+L_2}\sum_{L^{\prime }=|L-L_3|}^{L+L_3} \\
\times \Big[ (2L_1+1)(2L_2+1)(2L_3+1)(2L+1)(2L^{\prime }+1)\Big]^{1/2}C_{L_2}C_{L_3} \\
\times \Bigg\{ f_{1a}f_{2a}C_{1a}C_{La}C_{Lc}B_{1}(L^{\prime},M+1) 
- f_{1b}f_{2d}C_{1b}C_{Lb}C_{Ld}B_{1}(L^{\prime },M-1) \\ 
+\delta (L^{\prime },0) \Big(
f_{1a}f_{2b}C_{1a}C_{La}C_{Lc}-f_{1b}f_{2c}C_{1b}C_{Lb}C_{Ld}\Big) \Bigg\} \\
\times \Big[ J(N_1-2,N_2,N_3;\omega _1,\omega _2,\omega_3;1,1;L_2,L_3) 
-J(N_1,N_2;N_3;\omega _1,\omega _2,\omega _3;-1,1;L_2,L_3) \\
-J(N_1-2,N_2+2,N_3;\omega _1,\omega _2,\omega _3;-1,1;L_2,L_3)\Big]. 
\end{multline}

The integrals $B_{1}(L,M)$ are defined in Eq. (A.21).

\end{appendix}

\newpage

\begin{appendix}
\section*{Appendix C: Evaluation of the kinetic energy contribution ${\bf I}_{\bf \varphi}\textbf{(1)}$}

\setcounter{equation}{0}

\renewcommand{\theequation}{C.\arabic{equation}}

The last kinetic energy integral is the generated by the operator $\hat T_{\varphi }$ of Eq. (14):  
\begin{multline}
I_{\varphi}(1) = \left\langle \phi (\mathbf{r}_1)\phi (\mathbf{r}_2)\phi (%
\mathbf{r}_3)r_{13}|-\frac{r_2}{r_1r_{12}}\frac{\sin {\theta _2}}{\sin {%
\theta _1}}\sin {(\varphi _1-\varphi _2)}\frac{\partial ^2}{\partial \varphi
_1\partial r_{12}}|\phi^{\prime } (\mathbf{r}_1)\phi^{\prime } (\mathbf{r}_2)
\phi^{\prime } (\mathbf{r}_3)r_{12}\right\rangle . 
\end{multline}

As it is well-known, the derivative of a spherical harmonic with $m_1^{\prime }=0$ with respect
to $\varphi _1$ vanishes and so the whole integral:
\begin{equation}
I_{\varphi }(1)=0, 
\end{equation}

\noindent while for $m_1^{\prime } \neq 0$ the derivative is:

\begin{equation}
\frac{\partial Y_{l_1^{\prime}}^{m_1^{\prime}}(\theta_1 ,\varphi_1 )}{\partial \varphi_1 }
=im_1^{\prime} Y_{l_1^{\prime}}^{m_1^{\prime}}(\theta_1 ,\varphi_1 ).
\end{equation}

Let us start with the function $\sin (\varphi_1-\varphi_2)$ and write it in exponential form: 

\begin{equation}
\sin (\varphi_1-\varphi_2)=\frac 1{2i}\left( e^{i\varphi _1}e^{-i\varphi _2}-e^{-i\varphi
_1}e^{i\varphi _2}\right) . 
\end{equation}

\noindent The following functions can be written as spherical harmonics with $l=1$ and 
$m=1,-1$:

\begin{equation}
\sin \theta _2e^{-i\varphi _2}=\sqrt{\frac{8\pi }3}Y_1^{-1}(\theta _2,\varphi
_2),\qquad \sin \theta _2e^{i\varphi _2}=-\sqrt{\frac{8\pi }3}Y_1^1(\theta
_2,\varphi _2).
\end{equation}

\noindent The products of spherical harmonics with same arguments can be expanded:

\begin{multline}
\sqrt{\frac{8\pi }3}Y_{l_2}^{m_{2*}}(\theta _2,\varphi _2)Y_1^{-1}(\theta
_2,\varphi _2)=\sqrt{\frac 23}%
\sum_{L_2=|l_2-1|}^{l_2+1}(2L_2+1)^{1/2}C_{L_2a}Y_{L_2}^{-1-m_2}(\theta
_2,\varphi _2), \\
\sqrt{\frac{8\pi }3}Y_{l_2}^{m_{2*}}(\theta _2,\varphi _2)Y_1^1(\theta _2,\varphi
_2)=\sqrt{\frac 23}%
\sum_{L_2=|l_2-1|}^{l_2+1}(2L_2+1)^{1/2}C_{L_2b}Y_{L_2}^{1-m_2}(\theta _2,\varphi
_2), \\  
\end{multline}

\noindent with

\begin{equation}
C_{L_2a}=C^{L_2}(1,-1;l_2,m_2),\qquad C_{L_2b}=C^{L_2}(1,1;l_2,m_2).
\end{equation}

\noindent Using the above derived expressions, the product of angular functions of Eq. (C.1) 
is:

\begin{multline}
\left( -Y_{l_1}^{m_{1*}}(\theta _1,\varphi _1)Y_{l_2}^{m_2}(\theta _2,\varphi _2)%
\frac{\sin {\theta _2}}{\sin {\theta _1}}\sin {(\varphi _1-\varphi _2)}\frac{%
\partial Y_{l_1^{\prime }}^{m_1^{\prime }}(\theta _1,\varphi _1)}{\partial \varphi
_1}Y_{l_2^{\prime }}^{m_2^{\prime }}(\theta _2,\varphi _2)\right) \\ 
\times \Big( Y_{l_3}^{m_{3*}}(\theta _3,\varphi _3)Y_{l_3}^{m_3}(\theta
_3,\varphi _3)\Big) = \\
\frac 1{(4\pi )^{1/2}}\frac{m_1^{\prime}}2\sqrt{\frac 23}\sum_{L_3=|l_3^{\prime
}-l_3|}^{l_3^{\prime }+l_3}\sum_{L_2=|l_2-1|}^{l_2+1}\Big[
(2L_2+1)(2L_3+1)\Big] ^{1/2}C_{L_3}Y_{L_3}^{M_3}(\theta _3,\varphi _3) \\
\times \Bigg\{ C_{L_2a}Y_{L_2}^{-1-m_2}(\theta _1,\varphi _1)Y_{l_1^{\prime
}}^{m_1}(\theta _1,\varphi _1)\frac{e^{i\varphi _1}}{\sin \theta _1}Y_{l_1^{\prime
}}^{m_1^{\prime }}(\theta _1,\varphi _1)Y_{l_2^{\prime }}^{m_2^{\prime
}}(\theta _2,\varphi _2) \\
 +C_{L_2b}Y_{L_2}^{1-m_2}(\theta _1,\varphi _1)Y_{l_1^{\prime
}}^{m_1}(\theta _1,\varphi _1)\frac{e^{-i\varphi _1}}{\sin \theta _1}%
Y_{l_1^{\prime }}^{m_1^{\prime }}(\theta _1,\varphi _1)Y_{l_2^{\prime
}}^{m_2^{\prime }}(\theta _2,\varphi _2)\Bigg\} ,
\end{multline}

\noindent with $C_{L_3}=C^{L_3}(l_3^{\prime },m_3^{\prime };l_3,m_3)$. Using the
recursion relation containing the inverse sine function Eq. (1) over $%
Y_{l_1^{\prime }}^{m_1^{\prime }}(\theta _1,\varphi _1)$ (note that this relation can be
always used directly over $Y_{l_1^{\prime }}^{m_1^{\prime }}(\theta _1,\varphi
_1)$ since by definition $m_1^{\prime }\neq 0$, for $m_1^{\prime }=0$, the
derivative of a spherical harmonic is zero, see Eq. (C.1)) and the factors
defined as:

\begin{equation}
f_{1a}=\Big[(l_1^{\prime }-m_1^{\prime }+2)(l_1^{\prime }-m_1^{\prime
}+1)\Big]^{1/2},\qquad f_{1b}=\Big[(l_1^{\prime }+m_1^{\prime }+2)(l_1^{\prime
}+m_1^{\prime }+1)\Big]^{1/2}. 
\end{equation}

\noindent Eq. (C.8) can be written:

\begin{multline}
(-1)^{m_2^{\prime }}\frac 1{(4\pi )^{1/2}}\frac 14\sqrt{\frac 23}\left[ 
\frac{2l_1^{\prime }+1}{2l_1^{\prime }+3}\right]
^{1/2}\sum_{L_3=|l_3^{\prime }-l_3|}^{l_3^{\prime
}+l_3}\sum_{L_2=|l_2-1|}^{l_2+1}\Big[ (2L_2+1)(2L_3+1)\Big] ^{1/2} \\
\times C_{L_3}Y_{L_3}^{M_3}(\theta _3,\varphi _3)\Bigg\{ C_{L_2a}Y_{l_2^{\prime
}}^{m_2^{\prime }}(\theta _2,\varphi _2)Y_{L_2}^{-1-m_2}(\theta _2,\varphi
_2) \\
\times \Big( f_{1a}Y_{l_1^{\prime }}^{m_1}(\theta _1,\varphi _1) 
Y_{l_1^{\prime }+1}^{m_1^{\prime }-1}(\theta _1,\varphi_1)e^{2\varphi_1} 
+ f_{1b}Y_{l_1^{\prime }}^{m_1}(\theta _1,\varphi _1)
Y_{l_1^{\prime}+1}^{m_1^{\prime }+1}(\theta _1,\varphi _1)\Big) \\
+C_{L_2b}Y_{l_2^{\prime }}^{m_2^{\prime }}(\theta _2,\varphi_2)
Y_{L_2}^{1-m_2}(\theta _2,\varphi _2) \\
\times \Big( f_{1a}Y_{l_1^{\prime }}^{m_1}(\theta _1,\varphi
_1)Y_{l_1^{\prime }+1}^{m_1^{\prime }-1}(\theta _1,\varphi
_1)+f_{1b}Y_{l_1^{\prime }}^{m_1}(\theta _1,\varphi _1)
Y_{l_1^{\prime }+1}^{m_1^{\prime }+1}(\theta _1,\varphi _1) e^{-2\varphi_1}
\Big) \Bigg\}. 
\end{multline}

Combining $Y_{l_2^{\prime }}^{m_2^{\prime }}(\theta _2,\varphi
_2)Y_{L_2}^{-1-m_2}(\theta _2,\varphi _2)$:

\begin{eqnarray}
Y_{l_2^{\prime }}^{m_2^{\prime }}(\theta _2,\varphi _2)Y_{L_2}^{-1-m_2}(\theta
_2,\varphi _2) &=&(-1)^{m_2^{\prime }}\sum_{L_2^{\prime }=|L_2-l_2^{\prime
}|}^{L_2+l_2^{\prime }}\frac{(2L_2^{\prime }+1)^{1/2}}{(4\pi )^{1/2}}%
C_{L_2^{\prime }a}Y_{L_2^{\prime }}^{M_2-1}(\theta _2,\varphi _2) ,  \nonumber \\
Y_{l_2^{\prime }}^{m_2^{\prime }}(\theta _2,\varphi _2)Y_{L_2}^{1-m_2}(\theta
_2,\varphi _2) &=&(-1)^{m_2^{\prime }}\sum_{L_2^{\prime }=|L_2-l_2^{\prime
}|}^{L_2+l_2^{\prime }}\frac{(2L_2^{\prime }+1)^{1/2}}{(4\pi )^{1/2}}%
C_{L_2^{\prime }b}Y_{L_2^{\prime }}^{M_2+1}(\theta _2,\varphi _2) , 
\end{eqnarray}
with
\begin{equation}
C_{L_2^{\prime }a}=C^{L_2^{\prime }}(L_2,-1-m_2;l_2^{\prime },-m_2^{\prime
}),\qquad c_{L_2^{\prime }b}=C^{L_2^{\prime }}(L_2,1-m_2;l_2^{\prime
},-m_2^{\prime }) , 
\end{equation}

\noindent and $M_2= m_2^{\prime}-m_2$. Furthermore, combining the spherical harmonics with argument 1 and 
the ones with argument 3, Eq. (C.10) can be rewritten as:

\begin{multline}
(-1)^{m_2^{\prime }}\frac 1{(4\pi)^{3/2} }\frac 14\sqrt{\frac 23}\left[ \frac{%
2l_1^{\prime }+1}{2l_1^{\prime }+3}\right] ^{1/2}\sum_{L_3=|l_3^{\prime
}-l_3|}^{l_3^{\prime }+l_3}\sum_{L_2=|l_2-1|}^{l_2+1}\sum_{L_2^{\prime
}=|L_2-l_2^{\prime }|}^{L_2+l_2^{\prime }}\sum_{L_1=|l_1^{\prime
}+1-l_1|}^{l_1^{\prime }+1+l_1} \\
\times \Big[ (2L_1+1)(2L_2+1)(2L_3+1)(2L_2^{\prime }+1)\Big] ^{1/2} \\
\times \Bigg\{ C_{L_2a}C_{L_2^{\prime }a}c_{L_3}Y_{L_3}^{M_3}(\theta _3,\varphi
_3)Y_{L_2^{\prime }}^{M_2-1}(\theta _2,\varphi _2) \\
\times \Big( f_{1a}C_{L_1a}Y_{L_1}^{M_1-1}(\theta _1,\varphi _1)e^{2\varphi
_1}+f_{1b}C_{L_1b}Y_{L_1}^{M_1+1}(\theta _1,\varphi _1)\Big) \\
+C_{L_2b}C_{L_2^{\prime }b}c_{L_3}Y_{L_3}^{M_3}(\theta _3,\varphi
_3)Y_{L_2^{\prime }}^{M_2+1}(\theta _2,\varphi _2) \\
\times  \Big( f_{1a}C_{L_1a}Y_{L_1}^{M_1-1}(\theta _1,\varphi
_1)+f_{1b}C_{L_1b}Y_{L_1}^{M_1+1}(\theta _1,\varphi _1)e^{-2\varphi _1}\Big)
\Bigg\} ,
\end{multline}
with
\begin{equation}
C_{L_1a}=C^{L_1}(l_1^{\prime }+1,m_1^{\prime }-1;l_1,m_1),\qquad
C_{L_1b}=C^{L_1}(l_1^{\prime }+1,m_1^{\prime }+1;l_1,m_1).
\end{equation}

After the rotations of the spherical harmonics of electrons 2 and 3, see Eq. (A.11), 
and writing the integral of the radial part in form of a $J$-integral:

\begin{multline}
I_{\varphi}(1)=(-1)^{m_2^{\prime }}(4\pi)^{1/2}\frac 14\sqrt{\frac 23}\left[ \frac{%
2l_1^{\prime }+1}{2l_1^{\prime }+3}\right] ^{1/2}\sum_{L_3=|l_3^{\prime
}-l_3|}^{l_3^{\prime }+l_3}\sum_{L_2=|l_2-1|}^{l_2+1}\sum_{L_2^{\prime
}=|L_2-l_2^{\prime }|}^{L_2+l_2^{\prime }}\sum_{L_1=|l_1^{\prime
}+1-l_1|}^{l_1^{\prime }+1+l_1} \\
\times \Big[ (2L_1+1)(2L_2+1)(2L_3+1)(2L_2^{\prime }+1)\Big]
^{1/2}C_{L_3} \Bigg\{ C_{L_2a}C_{L_2^{\prime }a} \\
\times \Big( f_{1a}C_{L_1a}\int_0^\pi \int_0^{2\pi }Y_{L_3}^{M_3}(\theta
_1,\varphi _1)Y_{L_1}^{M_1-1}(\theta _1,\varphi _1)Y_{L_2^{\prime
}}^{M_2-1}(\theta _1,\varphi _1)e^{2\varphi _1}\sin (\theta _1)d\theta _1d\varphi_1 \\
+f_{1b}C_{L_1b}\int_0^\pi \int_0^{2\pi }Y_{L_3}^{M_3}(\theta _1,\varphi
_1)Y_{L_1}^{M_1+1}(\theta _1,\varphi _1)Y_{L_2^{\prime }}^{M_2-1}(\theta
_1,\varphi _1)\sin (\theta _1)d\theta _1d\varphi_1 \Big) \\
+C_{L_2b}C_{L_2^{\prime }b}\Big( f_{1a}C_{L_1a}\int_0^\pi \int_0^{2\pi
}Y_{L_3}^{M_3}(\theta _1,\varphi _1)Y_{L_1}^{M_1-1}(\theta _1,\varphi
_1)Y_{L_2^{\prime }}^{M_2+1}(\theta _1,\varphi _1)\sin (\theta _1)d\theta _1d\varphi_1 \\
+ f_{1b}C_{L_1b}\int_0^\pi \int_0^{2\pi }Y_{L_3}^{M_3}(\theta
_1,\varphi_1)Y_{L_1}^{M_1+1}(\theta_1,\varphi_1)Y_{L_2^{\prime
}}^{M_2+1}(\theta_1,\varphi_1)e^{-2\varphi_1}\sin (\theta _1)d\theta _1d\varphi_1 
\Big) \Bigg\} \\
\times J(N_1-1,N_3,N_2+1;\omega _1,\omega _3,\omega
_2;1,-1;L_3,L_2^{\prime })
\end{multline}

\noindent taking the complex conjugate and linearizing again the product of spherical harmonics with argument 1:
\begin{multline}
I_{\varphi}(1)=(-1)^{m_2^{\prime }+M_3}\frac 14\sqrt{\frac 23}\left[ 
\frac{2l_1^{\prime }+1}{2l_1^{\prime }+3}\right]^{1/2}\delta (M_1+M_2+M_3,0) 
\sum_{L_3=|l_3^{\prime }-l_3|}^{l_3^{\prime
}+l_3}\sum_{L_2=|l_2-1|}^{l_2+1}\sum_{L_2^{\prime }=|L_2-l_2^{\prime
}|}^{L_2+l_2^{\prime }} \\
\times \sum_{L_1=|l_1^{\prime }+1-l_1|}^{l_1^{\prime }+1+l_1} 
\sum_{L_3^{\prime }=|L_3-L_2^{\prime }|}^{L_3+L_2^{\prime }}\Big[
(2L_1+1)(2L_2+1)(2L_3+1)(2L_2^{\prime }+1)(2L_3^{\prime }+1)\Big]^{1/2} C_{L_3}\\
\times \Bigg\{ C_{L_2a}C_{L_2^{\prime }a} 
\Bigg( f_{1a}C_{L_1a}C_{L_3^{\prime}a}\int_0^\pi \int_0^{2\pi
}Y_{L_1}^{M_1-1}(\theta _1,\varphi _1)Y_{L_3^{\prime }}^{M_2+M_3-1}(\theta
_1,\varphi _1)e^{2\varphi _1}\sin (\theta _1)d\theta _1d\varphi_1 \\
+f_{1b}C_{L_1b}C_{L_3^{\prime}b}\int_0^\pi \int_0^{2\pi }Y_{L_1}^{M_1+1}(\theta
_1,\varphi _1)Y_{L_3^{\prime }}^{M_2+M_3-1}(\theta _1,\varphi _1)
\sin (\theta _1)d\theta _1d\varphi_1 \Big)\\
+C_{L_2b}C_{2L^{\prime }b}\Big( f_{1a}C_{L_1a}C_{L_3^{\prime}a}\int_0^\pi
\int_0^{2\pi }Y_{L_1}^{M_1-1}(\theta _1,\varphi _1)Y_{L_3^{\prime
}}^{M_2+M_3+1}(\theta _1,\varphi _1)\sin (\theta _1)d\theta _1d\varphi_1 \\
+ f_{1b}C_{L_1b}C_{L_3^{\prime}b}\int_0^\pi \int_0^{2\pi
}Y_{L_1}^{M_1+1}(\theta _1,\phi _1)Y_{L_3^{\prime }}^{M_2+M_3+1}(\theta
_1,\varphi_1)e^{-2\varphi_1}\sin (\theta _1)d\theta _1d\varphi_1 \Big) \Bigg\} \\
\times J(N_1-1,N_3,N_2+1;\omega _1,\omega _3,\omega
_2;1,-1;L_3,L_2^{\prime }), 
\end{multline}

\noindent with
\begin{equation}
C_{L_3^{\prime}a}=C^{L_3^{\prime }}(L_2^{\prime },M_2+1;L_3,-M_3),\qquad
C_{L_3^{\prime}b}=C^{L_3^{\prime }}(L_2^{\prime },M_2-1;L_3,-M_3).
\end{equation}

\noindent Let us define the new auxiliary angular integral:

\begin{multline}
B_{2}(L_1,M_1+n;L_2,M_2+n)=\int_0^\pi \int_0^{2\pi
}Y_{L_1}^{M_1+n}(\theta ,\phi )Y_{L_2}^{M_2+n}(\theta ,\phi )e^{-2ni\varphi
}\sin (\theta )d\theta d\varphi \\
=\frac 12\Big[ (2L_1+1)(2L_2+1)\Big] ^{1/2}\left[ \frac{(L_1-M_1-n)!}{%
(L_1+M_1+n)!}\right]^{1/2} \left[ \frac{(L_2-M_2-n)!}{%
(L_2+M_2+n)!}\right]^{1/2} P(L_1,M_1+n;L_2,M_2+m),  
\end{multline}
 
\noindent and evaluate it using the algorithm developed by Wong \cite{Wong}, in which the overlap
integral over associated Legendre functions is:

\begin{multline}
 P(L_1,M_1+n;L_2,M_2+m) = 
\int_0^\pi P_{L_1}^{M_1}(\cos \theta )P_{L_2}^{M_2}(\cos \theta )\sin
(\theta )d\theta =\sum_{p_{1=0}}^{p_{1\max }}\sum_{p_{2=0}}^{p_{2\max
}}a_{L_{1,}M_1}^{p_1}a_{L_{2,}M_2}^{p_2} \\
\times \frac{\Gamma \left( \frac 12(L_1+L_2-M_1-M_2-2p_1-2p_2+1)\right)
\Gamma \left( \frac 12(M_1+M_2+2p_1+2p_2+2)\right) }{\Gamma \left( \frac
12(L_1+L_2+3)\right) }.
\end{multline}
\bigskip

\noindent $\Gamma $ are Gamma functions. $p_{\max }=[(L-M)/2]$ is the integral part of $%
(L-M)/2$. This formula is valid for $0\leq M_1\leq L_1$ and $0\leq M_2\leq
L_2$. For $M_1\leq L_1$ or $M_1\leq L_1$ the integral is zero, see Ref. \cite
{Wong}. If $M$ is negative the following formula Eq. (A.24) is used. 
For even $L_1+L_2-M_1-M_2$ the integral vanish. The coefficients are:

\bigskip
\begin{equation}
a_{L_{,}M}^p=\frac{(-1)^p(L+M)!}{2^{M+2p}(M+p)!p!(L-M-2p)!}.
\end{equation}

\newpage 
Finally, the integral is:

\begin{multline}
I_{\varphi}(1)=(-1)^{m_2^{\prime }+M_3}\frac 14\sqrt{\frac 23}\left[ 
\frac{2l_1^{\prime }+1}{2l_1^{\prime }+3}\right]^{1/2}\delta (M_1+M_2+M_3,0) 
\sum_{L_3=|l_3^{\prime }-l_3|}^{l_3^{\prime
}+l_3}\sum_{L_2=|l_2-1|}^{l_2+1}\sum_{L_2^{\prime }=|L_2-l_2^{\prime
}|}^{L_2+l_2^{\prime }} \\ 
\times \sum_{L_1=|l_1^{\prime }+1-l_1|}^{l_1^{\prime }+1+l_1}
\sum_{L_3^{\prime }=|L_3-L_2^{\prime }|}^{L_3+L_2^{\prime }} 
\Big[(2L_1+1)(2L_2+1)(2L_3+1)(2L_2^{\prime }+1)(2L_3^{\prime }+1)\Big]^{1/2} C_{L_3}\\ 
\times \Bigg\{ C_{L_2a}C_{L_2^{\prime }a} 
\Big( f_{1a}C_{L_1a}C_{L_3^{\prime}a}B_{2}(L_1,M_1-1;L_3^{\prime },M_2+M_3-1) 
+ (-1)^{M_1+1}\; \delta (L_1,L_3^{\prime }) f_{1b}C_{L_1b}C_{L_3^{\prime}b} \Big) \\
+C_{L_2b}C_{L_2^{\prime }b}\Big( 
f_{1b}C_{L_1b}C_{L_3^{\prime}b}B_{2}(L_1,M_1+1;L_2^{\prime },M_2+M_3+1)
+ (-1)^{M_1-1}\;\delta(L_1,L_3^{\prime }) f_{1a}C_{L_1a}C_{L_3^{\prime}a}  
\Big) \Bigg\} \\
\times J(N_1-1,N_3,N_2+1;\omega _1,\omega _3,\omega_2;1,-1;L_3,L_2^{\prime })
\end{multline}

\end{appendix}

\newpage

\textbf{Figure 1:} Definition and rotation of the coordinates of two
electrons in an atomic center.

\begin{center}
\begin{figure}[th]
\includegraphics[scale=0.90,angle=-90]{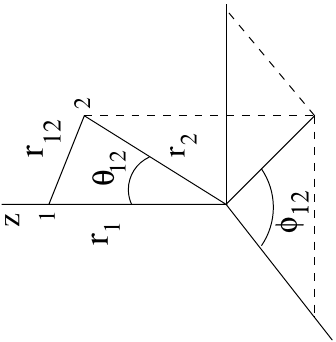}
\end{figure}
\end{center}

\end{document}